\documentclass[12pt,prb,superscriptaddress,preprintnumbers,amsmath,amssymb]{revtex4}

\usepackage{graphicx} 
\usepackage{dcolumn} 

\def\bG{{\bf G}}

\def\bk{{\bf k}}

\def\bp{{\bf p}}
\def\bq{{\bf q}}

\def\bG{{\bf G}}
\def\bQ{{\bf Q}}

\def\b0{{\bf 0}}
\def\btau{{\boldsymbol\tau}}

\def\Re{{\rm Re}}
\def\Im{{\rm Im}}

\def\alf{\alpha}
\def\eps{\epsilon}

\def\Lam{\Lambda}
\def\om{\omega}

\def\sg{\sigma}
\def\Sg{\Sigma}

\def\sgn{\rm sgn}

\begin{document}

\title{Fluctuation effects at the onset of $\bf 2k_F$ density wave order \\
with two pairs of hot spots in two-dimensional metals}

\author{J\'achym S\'ykora}
\affiliation{Max Planck Institute for Solid State Research,
 D-70569 Stuttgart, Germany}
\author{Walter Metzner}
\affiliation{Max Planck Institute for Solid State Research,
 D-70569 Stuttgart, Germany}

\date{\today}

\begin{abstract}
We analyze quantum fluctuation effects at the onset of charge or spin density wave order in two-dimensional metals with an incommensurate nesting ($2k_F$) wave vector connecting two pairs of hot spots on the Fermi surface. We first compute the momentum and frequency dependence of the fermion self-energy near the hot spots to leading order in a perturbation expansion (one loop). Non-Fermi liquid behavior with a linear (in energy) quasi-particle decay rate and a logarithmically vanishing quasi-particle weight is obtained. The momentum dependence of the self-energy entails only finite renormalizations of the Fermi velocity and the Fermi surface curvature at the hot spots. The perturbative one-loop result is not self-consistent and casts doubt on the stability of the $2k_F$ quantum critical point. We construct a self-consistent solution of the one-loop equations with self-energy feedback, where the quantum critical point is stabilized rather than being destroyed by fluctuations, while the non-Fermi liquid behavior as found in the perturbative one-loop calculation is confirmed.
\end{abstract}

\maketitle


\section{Introduction}

Quantum critical fluctuations at the onset of charge or spin density wave order in two-dimensional metals destroy Fermi liquid behavior and lead to unconventional dependences on temperature and other control parameters. \cite{loehneysen07}
The theory of such systems is difficult due to a complex interplay of the critical order parameter fluctuations and the gapless excitations of the electrons. A purely bosonic order parameter theory as developed by Hertz \cite{hertz76} and Millis \cite{millis93} may describe some features, but it is not generally applicable since the electronic excitations lead to singular interactions of the order parameter fluctuations. \cite{abanov04} Secondary instabilities generated by the fluctuations, especially pairing instabilities, are almost unavoidable and further complicate the analysis.

The most intensively studied case is the quantum critical point at the onset of N\'eel-type antiferromagnetic order. \cite{abanov03} Non-Fermi liquid behavior is obtained already at leading order in perturbation theory, but new features appear at higher orders. \cite{metlitski10_af1} Impressive analytical \cite{lee18} and numerical \cite{berg12} efforts have led to substantial progress, but a complete theory of this important universality class has not yet been constructed.

A special and particularly intricate situation arises when the density wave vector is a {\em nesting vector}\/ of the Fermi surface, that is, when it connects Fermi points with antiparallel Fermi velocities. \cite{footnote_perfnest}
Charge and spin correlation functions exhibit a singularity at such wave vectors due to an enhanced phase space for low-energy particle-hole excitations. In inversion symmetric crystals with  a valence band dispersion $\eps_\bk$, nesting vectors $\bQ$ are determined by the condition $\eps_{(\bQ+\bG)/2} = \eps_F$, where $\bG$ is zero or a reciprocal lattice vector, and $\eps_F$ is the Fermi energy. Since nesting vectors in isotropic continuum systems are determined by the Fermi surface radius $k_F$ via the simple relation $|\bQ| = 2k_F$, we refer to nesting vectors also as ``$2k_F$-vectors''.

Nesting singularities are particularly pronounced in metals with reduced spatial dimensionality. Charge and spin susceptibilities in low-dimensional metals exhibit peaks at nesting vectors, so that these vectors are natural charge or spin density wave vectors. For example, the ground state of the two-dimensional Hubbard model undergoes a continuous quantum phase transition into a spin-density wave state with a wave vector satisfying the nesting condition, at least within mean-field theory. \cite{schulz90,igoshev10}
Also $d$-wave bond charge order generated by antiferromagnetic fluctuations in spin-fermion models for cuprates \cite{metlitski10_af2,sachdev13} occurs naturally with nesting wave vectors. \cite{holder12,punk15}

Quantum criticality in a two-dimensional metal at the onset of density wave order with a nesting wave vector was first analyzed by Altshuler {\em et al.} \cite{altshuler95}
They obtained non-Fermi liquid power laws for the electron self-energy and the susceptibility for the special case where the wave vector $\bQ$ is not only a nesting vector, but also half a reciprocal lattice vector.
Later it was shown that due to additional umklapp processes Landau quasi-particles actually survive, albeit with an enhanced decay rate, for the particular case where $\bQ$ is the antiferromagnetic wave vector $(\pi,\pi)$. \cite{bergeron12,wang13}
For generic, that is, incommensurate \cite{footnote_incomm} nesting vectors instead, Altshuler {\em et al.} \cite{altshuler95} found strong infrared divergences in two-loop contributions to the susceptibility and concluded that the order-parameter fluctuations destroy the quantum critical point, replacing the second-order by a first-order phase transition.

The analysis of non-Fermi liquid behavior at the onset of charge and spin-density wave order with an incommensurate nesting vector in two-dimensional metals was initiated more recently. The fermion self-energy was calculated to first order (one loop) in the order parameter fluctuation propagator. \cite{holder14} The fluctuation propagator was computed from a one-loop approximation with bare particle-hole bubbles. A breakdown of Landau Fermi liquid theory was obtained at {\em hot spots}\/ on the Fermi surface, that is, Fermi points connected by the ordering wave vector $\bQ$. Two qualitatively distinct cases arise according to the number of hot spots connected by $\bQ$. 

In the simplest case, $\bQ$ connects only a single pair of hot spots, and points typically in a high symmetry direction (axial or diagonal). The frequency dependence of the one-loop fermion self-energy at the hot spots follows a power law with an exponent $\frac{2}{3}$ in this case. \cite{holder14} The decay rate of fermionic excitations is thus larger than their excitation energy in the low energy limit. The momentum dependence yields a singular renormalization of the Fermi velocity and a flattening of the Fermi surface near the hot spots. \cite{sykora18} An important issue is the self-consistency of these results, or whether the non-Fermi liquid self-energy destroys the $2k_F$ quantum critical point, since it has a tendency to wipe out the $2k_F$ peak in the density wave susceptibility. It was shown that the $2k_F$ quantum critical point can survive only with the assistance of a sufficiently strong Fermi surface flattening around the hot spots. \cite{sykora18} This scenario was subsequently supported by a systematic $\eps$-expansion around the critical dimension $d_c = 5/2$, at least to leading order in $\eps$. \cite{halbinger19}

In the second case, the density wave vector $\bQ$ connects two pairs of hot spots. The Fermi velocities are collinear within each pair, but not between the pairs. In Fig.~1 we show various possible geometries, with Fermi surfaces and hot spot pairs on the left, and the lines of all nesting vectors $\bQ$ satisfying the condition $\eps_{(\bQ+\bG)/2} = \eps_F$ on the right. These lines are obtained by backfolding the line consisting of all doubled Fermi wave vectors $2\bk_F$ into the first Brillouin zone. Wave vectors $\bQ$ connecting two pairs of hot spots correspond to crossing points of the lines of nesting vectors.
The geometry in the top row of Fig.~1 is realized by the spin-density wave quantum critical point in the two-dimensional Hubbard model, while the $d$-wave bond charge order instability mentioned above provides an example for the geometry in the second row.
For two hot spot pairs, the imaginary part of the one-loop self-energy at the hot spots was found to be a linear function of frequency, with a distinct prefactor for positive and negative frequencies. \cite{holder14} Hence, the decay rate of fermionic excitations is proportional to their excitation energy, implying a marginal breakdown of Fermi liquid theory.
\begin{figure}
\centering
\includegraphics[width=4cm]{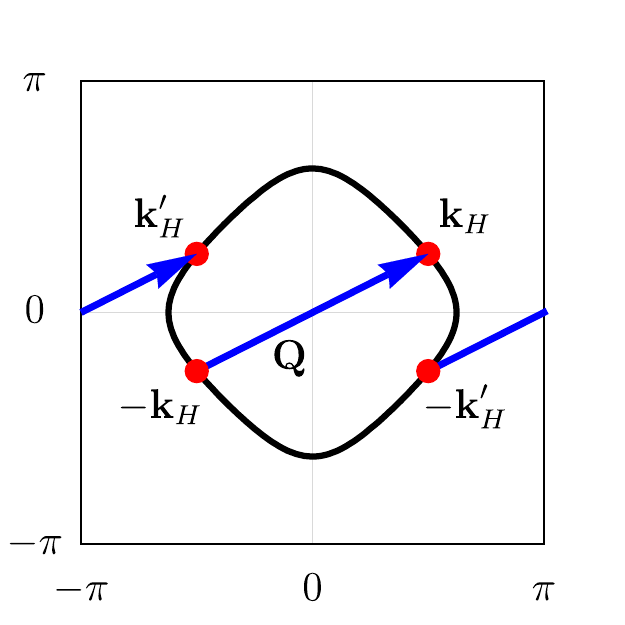} \hskip 5mm
\includegraphics[width=4cm]{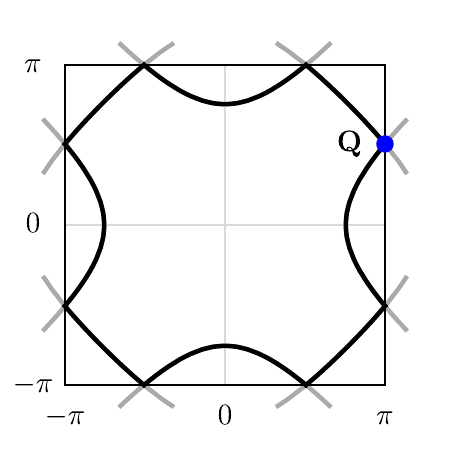} \\
\includegraphics[width=4cm]{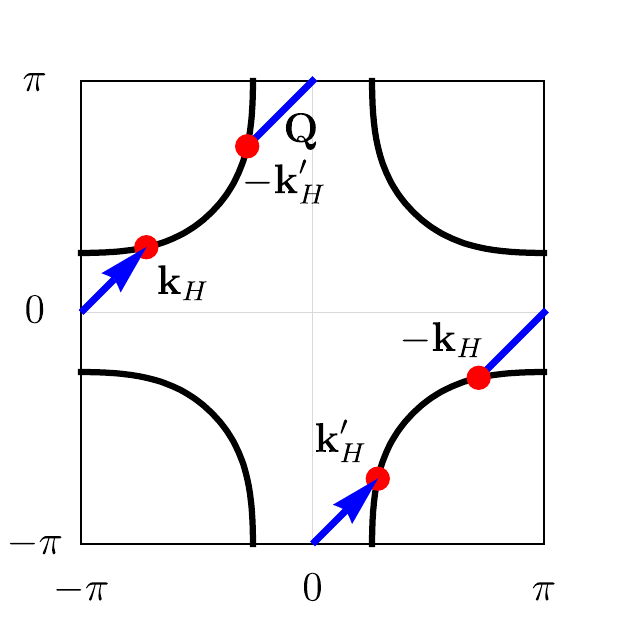} \hskip 5mm
\includegraphics[width=4cm]{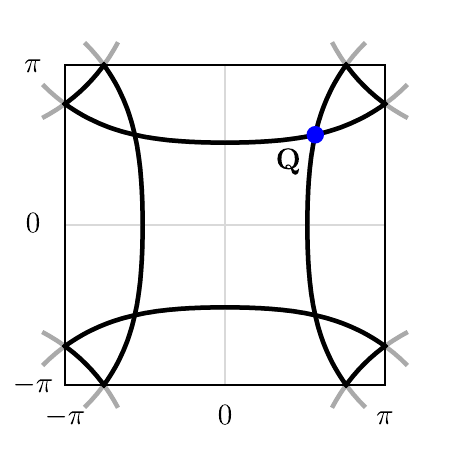} \\
\includegraphics[width=4cm]{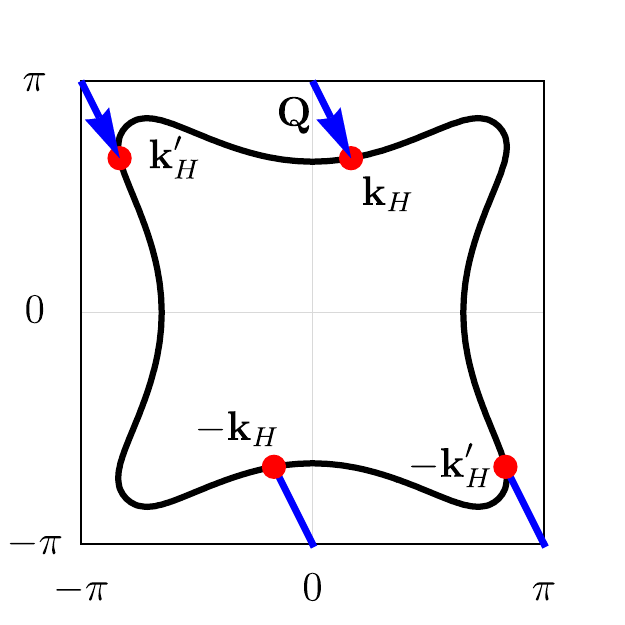} \hskip 5mm
\includegraphics[width=4cm]{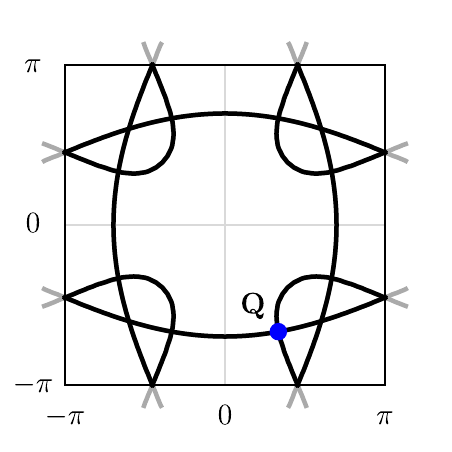}
\caption{Left: Fermi surfaces with two pairs of hot spots connected by a density wave
 vector $\bQ$. Right: Lines of nesting vectors and their crossing point corresponding
 to $\bQ$ in the left panel.
 Top: $\bQ = (\pi,Q)$ with an incommensurate $y$-component $Q$.
 Center: Diagonal wave vector $\bQ = (Q,Q)$ with incommensurate $Q$.
 Bottom: Wave vector $\bQ$ with two distinct incommensurate components.}
\end{figure}

In this work we extend the analysis of fluctuation effects at the onset of $2k_F$ density wave order with two pairs of hot spots. We compute the frequency dependence of the one-loop self-energy on the imaginary frequency axis and show that the results are consistent with the real frequency self-energy obtained previously. \cite{holder14}
We also compute the momentum dependence near the hot spots. Unlike the case of a single hot spot pair, \cite{sykora18} the momentum dependence of the self-energy entails only a finite renormalization of the Fermi velocity, and the renormalized Fermi surface maintains a finite curvature at the hot spots.

The perturbative one-loop results are not self-consistent. Incorporating the one-loop self-energy in the fermion propagator can lead to a shift of the peak in the susceptibility away from the nesting vector, which spoils the $2k_F$ quantum critical point.
To include the self-energy feedback self-consistently, we explore two distinct routes.
First, we compute the fluctuation propagator and the fermion self-energy from a renormalization group flow. Due to a peculiar cutoff dependence of the fermion self-energy, the self-energy feedback is incomplete in this approach and the shift of the susceptibility peak persists. Second, we obtain a fully self-consistent solution by solving the coupled integral equations for the particle-hole bubble and the fermion self-energy with self-energy feedback. Here the peak at the nesting vector survives.

The paper is structured as follows. In Sec.~II we derive the order parameter susceptibility and effective interaction at the quantum critical point in one-loop approximation with bare fermion propagators. In Sec.~III we evaluate the frequency and momentum dependence of the one-loop fermion self-energy near the hot spots. In Sec.~IV we show that the perturbative one-loop results are not self-consistent. In Sec.~V we discuss our renormalization group analysis of the quantum critical point, and in Sec.~VI we present the self-consistent solution of the one-loop equations with self-energy feedback. A conclusion in Sec.~VII closes the article.


\section{RPA susceptibility and effective interaction}

We consider a one-band system of interacting fermions with a bare dispersion relation $\eps_\bk$. Our calculations are based on the standard quantum many-body formalism with an imaginary frequency representation of dynamical quantities.\cite{negele87}
The bare fermion propagator is given by
\begin{equation}
 G_0(\bk,ik_0) = \frac{1}{ik_0 - \xi_\bk} \\ ,
\end{equation}
where $k_0$ is the frequency variable, and $\xi_\bk = \eps_\bk - \mu$ is the single-particle energy relative to the chemical potential $\mu$. At zero temperature the Matsubara frequency $k_0$ is a continuous variable.

We assume that, in mean-field theory, the system undergoes a charge or spin density-wave instability with an incommensurate and nested ($2k_F$) modulation vector $\bQ$ at a QCP which can be reached by tuning a suitable parameter such as electron density or interaction strength.
Approaching the QCP from the normal metallic regime, the instability is signalled by a diverging RPA susceptibility
\begin{equation} \label{chi}
 \chi(\bq,iq_0) = \frac{\chi_0(\bq,iq_0)}{1 + g \chi_0(\bq,iq_0)} \, ,
\end{equation}
where $g < 0$ is the coupling constant parametrizing the bare interaction in the instability channel, and $\chi_0$ is the bare susceptibility
\begin{equation} \label{chi_0}
 \chi_0(\bq,iq_0) =
 - N \int \frac{d^2\bk}{(2\pi)^2} \int \frac{dk_0}{2\pi} \,
 f_{\bk-\bq/2}^2 \, G_0(\bk,ik_0) \, G_0(\bk-\bq,ik_0-iq_0) \, .
\end{equation}
$N$ is the spin-multiplicity ($N=2$ for spin-$\frac{1}{2}$ fermions), and $f_\bp$ is a form factor related to the internal structure of the density-wave order parameter. For an order parameter with $s$-wave symmetry, $f_\bp$ is symmetric under rotations and reflections. In the following we assume $f_\bp = 1$ for definiteness. A generalization to form factors with other symmetries such as $d$-wave symmetry is straightforward.
Eq.~(\ref{chi}) holds also for the spin susceptibility in the normal (symmetric) phase of spin-rotation invariant systems, where all components of the spin susceptibility are equal.

%
\begin{figure}
\centering
\includegraphics[width=5cm]{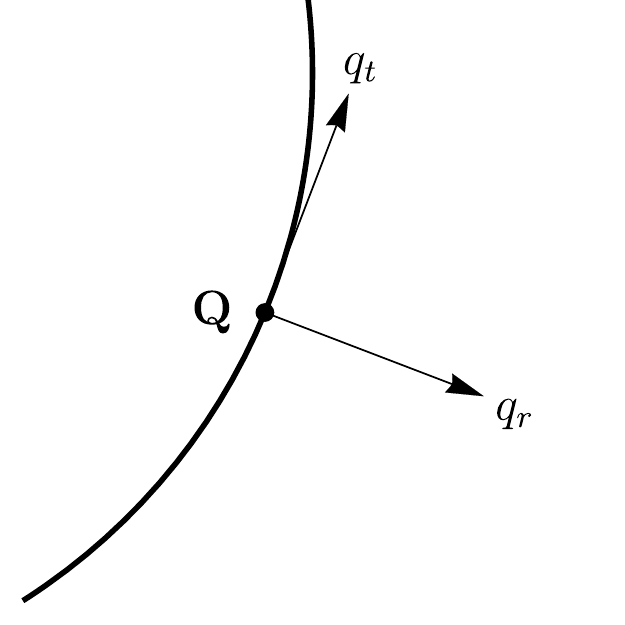}
\includegraphics[width=5cm]{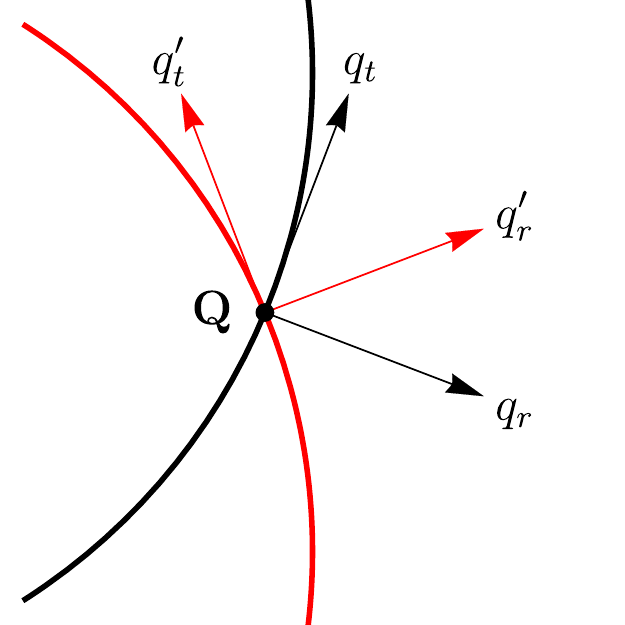}
\caption{Left: Normal and tangential momentum coordinates for momenta near a density 
 wave vector $\bQ$ on the $2k_F$ line. Right: Normal and tangential coordinates for
 momenta near $\bQ$ at a crossing point of two $2k_F$-lines.}
\end{figure}

The bare susceptibility $\chi_0(\bq,iq_0)$ exhibits a square-root singularity on the $2k_F$ line defined by all $2k_F$ vectors in $\bq$-space.
To parametrize momenta near a $2k_F$ momentum $\bQ$, we introduce relative coordinates normal and tangential to the $2k_F$ line at $\bQ$, which we denote by $q_r$ and $q_t$, respectively (see Fig.~2).
For momenta near $\bQ$ and low frequencies, the bare susceptibility can be expanded as \cite{altshuler95,holder14}
\begin{equation}
 \chi_0(\bq,iq_0) =
 \chi_0(\bQ,0) - N \big[ a \, h(e_\bq,q_0) - b \, e_\bq - c \, q_t \big] \, ,
\end{equation}
where $a$ and $b$ are positive constants, $c$ is a real constant, and
\begin{equation}
 h(e_\bq,q_0) = \sqrt{e_\bq + iq_0} + \sqrt{e_\bq - iq_0} =
 \sqrt{2} \sqrt{\sqrt{e_\bq^2 + q_0^2} + e_\bq} \, .
\end{equation}
The energy-momentum relation $e_\bq$ is given by
\begin{equation} \label{e_q}
e_\bq = v_F q_r + \frac{q_t^2}{4m} \, ,
\end{equation}
where $v_F$ is the Fermi velocity at the hot spots on the Fermi surface $\pm\bk_H$ connected by $\bQ$, and $m$ parametrizes the Fermi surface curvature at these points ($mv_F$ is the radius of curvature).
Hence, $e_\bq/v_F$ is the oriented distance of $\bq$ from the $2k_F$-line.
The prefactor of the square-root term is determined by the Fermi velocity and curvature near $\pm \bk_H$ as
\begin{equation}
 a = \frac{\sqrt{m}}{4\pi v_F} \, ,
\end{equation}
while the other constants $b$ and $c$ receive contributions from everywhere. For fermions with a parabolic dispersion in the continuum and a constant form factor, $b$ and $c$ vanish, as can be seen from Stern's exact analytic formula for $\chi_0$ in this case. \cite{stern67} The constant $c$ vanishes also at high symmetry points for lattice fermions, especially if $\bk_H$ points along an axis or a diagonal in the Brillouin zone. In this case a correction of order $q_t^2$ may become relevant. \cite{sykora18}

Throughout this article we consider the case where the ordering wave vector $\bQ$ connects {\em two}\/ pairs of hot spots $\pm\bk_H$ and $\pm\bk'_H$ on the Fermi surface. The wave vector $\bQ$ is then a crossing point of two $2k_F$ lines, as shown in Fig.~1 (right panel).
Let $q_r$ and $q_t$ be the normal and tangential coordinates relative to the first of the $2k_F$ lines as introduced above. The normal and tangential coordinates relative to the second $2k_F$ line, $q'_r$ and $q_t'$, respectively, are related to the former by
\begin{eqnarray}
 q_r' &=& q_r \cos\theta + q_t \sin\theta \, , \nonumber \\
 q_t' &=& q_t \cos\theta - q_r \sin\theta \, ,
\end{eqnarray}
where $\theta$ is the angle between the Fermi surface normal vectors at $\bk_H$ and $\bk'_H$, which is also the angle between the two $2k_F$ lines crossing at $\bQ$ (see Fig.~2). The $2k_F$ singularities on these two $2k_F$ lines need to be added, such that \cite{holder14}
\begin{equation} \label{chi_0_exp}
 \chi_0(\bq,iq_0) =
 \chi_0(\bQ,0) - N \left[
 a \, h(e_\bq,q_0) + a' h(e'_\bq,q_0) - b \, e_\bq - c \, q_t
 \right] \, ,
\end{equation}
where $e_\bq$ and $q_t$ are defined relative to the first $2k_F$ line, while
$e'_\bq = v'_F q'_r + \frac{1}{4m'} {q'_t}^2$ is defined relative to the second $2k_F$-line. The last two terms in Eq.~(\ref{chi_0_exp}) describe the leading regular momentum dependence near $\bQ$. We have chosen the variables $e_\bq$ and $q_t$ to parametrize this dependence. \cite{footnote_chi_0}
The hot spots are often related by a point group symmetry of the lattice, as in the first two examples in Fig.~1, so that $v_F = v'_F$, $m = m'$, and $a = a'$. The bare susceptibility exhibits a peak at $\bQ$ if and only if $c \, \theta > 0$. In the following we assume that this is the case.

At the QCP one has $g \chi_0(\bQ,0) = - 1$, so that the RPA susceptibility assumes the singular form
\begin{equation}
 \chi(\bq,iq_0) = - \frac{g^{-1} \chi_0(\bQ,0)}
 {N \big[
 a \, h(e_\bq,q_0) + a' h(e'_\bq,q_0) - b \, e_\bq - c \, q_t \big] } \, . 
\end{equation}

To deal with the critical order parameter fluctuations, the perturbation expansion has to be organized in powers of a dynamical effective interaction. This arises naturally as a boson propagator by decoupling the bare interaction in the instability channel via a Hubbard-Stratonovich transformation.\cite{hertz76,millis93} Alternatively it can be obtained by an RPA resummation of particle-hole bubbles or ladders.
In the simplest case of a charge-density wave instability in a spinless fermion system, the RPA effective interaction can be written as
\begin{equation} \label{D_rpa}
 D(\bq,iq_0) = \frac{g}{1 + g \chi_0(\bq,iq_0)} \, .
\end{equation}
For a charge-density wave instability in a spin-$\frac{1}{2}$ fermion system, the effective interaction has the diagonal spin structure
\begin{equation}
 D_{\sg'_1\sg'_2\sg_1\sg_2}(\bq,iq_0) =
 \delta_{\sg_1\sg'_1} \delta_{\sg_2\sg'_2} D(\bq,iq_0) \, ,
\end{equation}
where $\sg_1,\sg_2$ ($\sg'_1,\sg'_2$) are the spin-indices of the ingoing (outgoing) fermions.
For a spin-density wave, the effective interaction acquires a non-diagonal spin structure. In a spin-rotation invariant system of spin-$\frac{1}{2}$ fermions, it can be written as
\begin{equation} \label{D_spin}
 D_{\sg'_1\sg'_2\sg_1\sg_2}(\bq,iq_0) =
 \btau_{\sg_1\sg'_1} \cdot \btau_{\sg_2\sg'_2} \, D(\bq,iq_0) \, ,
\end{equation}
where $\btau = (\tau^x,\tau^y,\tau^z)$ is the vector formed by the three Pauli matrices $\tau^x,\tau^y,\tau^z$. \cite{footnote3}

Expanding the bare susceptibility as in Eq.~(\ref{chi_0_exp}), the effective interaction at the QCP assumes the asymptotic form
\begin{equation} \label{Dcrit}
 D(\bq,iq_0) =
 - \, \frac{1}{N \big[
 a \, h(e_\bq,q_0) + a' h(e'_\bq,q_0) - b \, e_\bq - c \, q_t \big] } \, . 
\end{equation}
It thus features the same singularity as the RPA susceptibility.
In case that the coupling $g$ has a (regular) momentum dependence for $\bq$ near $\bQ$, the coefficients $b$ and $c$ are not determined by $\chi_0(\bq,0)$ only, but receive additional contributions from the expansion of $g^{-1}(\bq)$ around $\bQ$.


\section{One-loop fermion self-energy}

To leading order in the effective interaction, the fermion self-energy is given by the one-loop expression
\begin{equation} \label{Sigma1}
 \Sg(\bk,ik_0) = - M \int \frac{d^2\bq}{(2\pi)^2} \int \frac{dq_0}{2\pi} \,
 D(\bq,iq_0) \, G_0(\bk-\bq,ik_0-iq_0) \, ,
\end{equation}
with $M=1$ for a charge density and $M=3$ for a spin density instability.
We evaluate $\Sg(\bk,ik_0)$ for low frequencies $k_0$ and momenta $\bk$ near one of the hot spots, say $\bk_H$.
The dominant contributions come from momentum transfers $\bq$ near $\bQ$, such that $\bk-\bq$ is situated near the antipodal hot spot $-\bk_H$.

In our derivations we assume that the Fermi surface is convex at the hot spots.
Results for the case of a concave Fermi surface follow from a simple particle-hole transformation. Using normal and tangential coordinates for $\bk$ near $\bk_H$, we expand the dispersion relation to leading order as
$\xi_\bk = v_F k_r + \frac{1}{2m} k_t^2$, and
$\xi_{\bk-\bq} = - v_F (k_r-q_r) + \frac{1}{2m}(k_t-q_t)^2$ for $\bq$ near $\bQ$.

Inserting Eq.~(\ref{Dcrit}) into Eq.~(\ref{Sigma1}), the one-loop self-energy is thus given by the integral
\begin{eqnarray} \label{Sigma2}
 \Sg(\bk,ik_0) &=& \frac{M}{N} \int \frac{dq_t}{2\pi} 
 \int \frac{dq_r}{2\pi} 
 \int \frac{dq_0}{2\pi} \, 
 \frac{1}{i(k_0-q_0) + v_F(k_r-q_r) - \frac{1}{2m}(k_t-q_t)^2} \nonumber \\
 & \times & \frac{1}
 {a \, h(e_\bq,q_0) + a' h(e'_\bq,q_0) - b \, e_\bq - c \, q_t} \, .
\end{eqnarray}
Note that we have not yet introduced an ultraviolet cutoff for the momentum integral. We will do so at a later stage of the evaluation, whenever needed.


\subsection{Frequency dependence at hot spot}

For finite frequencies ($k_0 \neq 0$) the self-energy is complex. The imaginary part of the integral in Eq.~(\ref{Sigma2}) diverges logarithmically for large momenta, such that an ultraviolet cutoff is needed. In the low-frequency limit, the imaginary part of the self-energy at a hot spot can be computed analytically (see Appendix A), and behaves as
\begin{equation} \label{ImSigma1}
 \Im\Sg(\bk_H,ik_0) = 
 - \frac{M k_0}{\pi N \alf} \, \ln \frac{\alf\Lam}{|k_0|} + {\cal O}(k_0) \, ,
\end{equation}
where 
\begin{equation}
 \alf = 8\pi v_F |c| \, ,
\end{equation}
and $\Lam$ is an ultraviolet momentum cutoff.
In the evaluation of $\Im\Sg(\bk_H,ik_0)$, a cutoff limiting the geometric mean of
$|e_{\bq}|$ and $q_t^2/4m$ turned out to be sufficient and convenient (see Appendix A). Other choices of a cutoff would merely amount to a different factor in the argument of the logarithm.

\begin{figure}
\centering
\includegraphics[width=10cm]{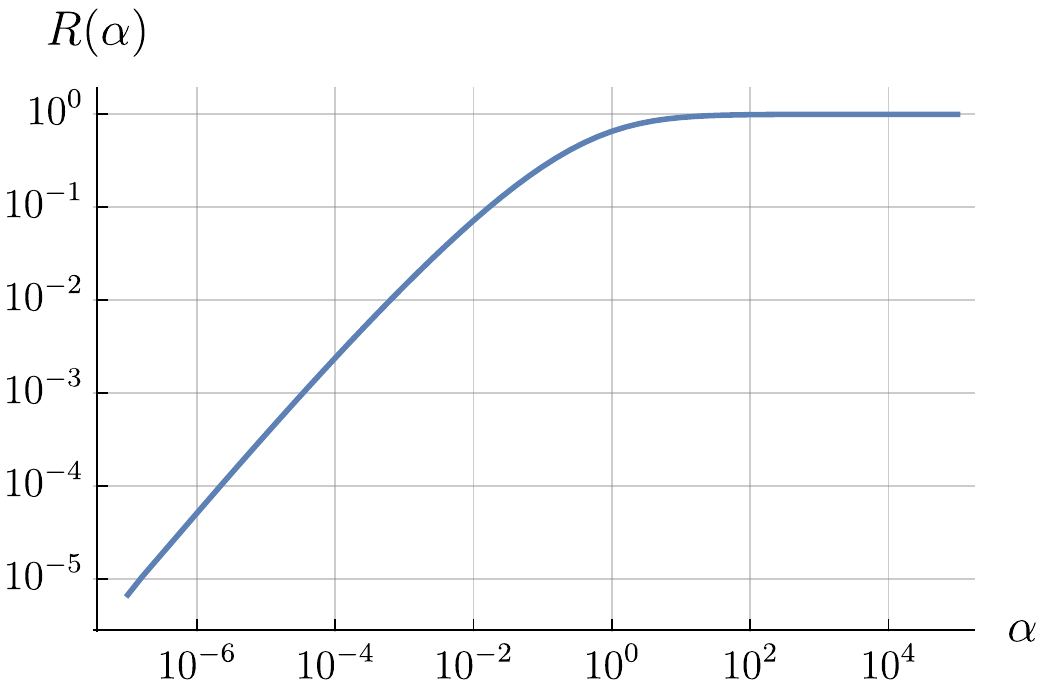}
\caption{The function $R(\alf)$.}
\label{fig:R}
\end{figure}

Subtracting the zero frequency constant, the real part of the self-energy is finite, that is, it does not require any ultraviolet cutoff. A simple rescaling of variables yields
\begin{equation} \label{ReSigma1}
 \Re\Sg(\bk_H,ik_0) - \Sg(\bk_H,0) =
 - \frac{M |k_0|}{\pi N \alf} \, R(\alf)
\end{equation}
for $k_0 \to 0$, where $R(\alf)$ is a real function depending only on the parameter $\alf$. The function $R(\alf)$ is given by a double integral which doesn't seem to be elementary (see Eq.~(\ref{R_alf}) in Appendix A). However, we could show analytically that $R(\alf) \to 1$ for $\alf \to \infty$. In Fig.~\ref{fig:R} we plot $R(\alf)$ as obtained from a numerical integration of Eq.~(\ref{R_alf}.

The leading low frequency behavior of the one-loop self-energy, Eqs.~(\ref{ImSigma1}) and (\ref{ReSigma1}), exhibits a remarkable cancellation. Although the tangential momentum dependence of $\xi_\bk$ and the corresponding Fermi surface curvature were crucial in deriving these results, the mass $m$, which determines the curvature radius, drops out. Moreover, the parameters characterizing the fermion dispersion near $\bk'_H$ do not appear in the asymptotic expressions.

In Ref.~\onlinecite{holder14} the one-loop self-energy was evaluated on the {\em real}\/ frequency axis. The result obtained in that paper can be written in the following form \cite{footnote4}
\begin{equation} \label{SigmaHM14}
 \Im\Sg(\bk_H,\om+i0^+) = - \frac{M}{N} C_{s_\om}(\alf) |\om| \, ,
\end{equation}
where $s_\om$ is the sign of $\om$, and
\begin{equation} \label{C_alf}
 C_{s_\om}(\alf) = \frac{s_\om}{\pi} 
 \int_0^1 \! d \tilde\om \int_0^{\infty}
 \frac{d \tilde\kappa}{\sqrt{\tilde\kappa}} \,
 \Im \left[ \sqrt{(2\tilde\om - 1) s_\om + i0^+ - \tilde\kappa} +
 \sqrt{s_\om - i0^+ - \tilde\kappa} + \alf \sqrt{\tilde\kappa} \right]^{-1} .
\end{equation}
For large $\alf$, these coefficients have the simple form 
$C_+ = \big( \frac{1}{2} - \frac{1}{\pi} \big) \frac{1}{\alf}$ and
$C_- = \big( \frac{1}{2} + \frac{1}{\pi} \big) \frac{1}{\alf}$.
We can relate these results for the one-loop self-energy to our results on the imaginary frequency axis by using the Kramers-Kronig-type relation for the self-energy in the complex frequency plane
\begin{equation} \label{KK}
 \Sigma(\bk_H,z) = - \frac{1}{\pi} \int_{-\infty}^{\infty} d\om \,
 \frac{\Im\Sg(\bk_H,\om+i0^+)}{z - \om} \, ,
\end{equation}
where $z$ is an arbitrary complex number (but not real), and we have dropped the momentum argument. Inserting the real-frequency behavior (\ref{SigmaHM14}), one finds
\begin{equation}
 \Re\Sg(\bk_H,ik_0) - \Sg(\bk_H,0) =
 \frac{M}{2N} \big[ C_+(\alf) - C_-(\alf) \big] |k_0| \, .
\end{equation}
For large $\alf$, this implies
$\Re\Sg(\bk_H,ik_0) - \Sg(\bk_H,0) = - \frac{M}{\pi N \alf} |k_0|$ in agreement with Eq.~(\ref{ReSigma1}).
For $z=ik_0$, the imaginary part of the integral in Eq.~(\ref{KK}) diverges logarithmically. Restricting the frequency integration by an ultraviolet cutoff $|\om| < \Lam_\om$ one obtains
\begin{equation} \label{ImSigma2}
 \Im\Sg(\bk_H,ik_0) = - \frac{M}{\pi N} \, 
 \big[ C_+(\alf) + C_-(\alf) \big] \, k_0 \, \ln \frac{\Lam_\om}{|k_0|} +
 {\cal O}(k_0) \, .
\end{equation}
This is consistent with Eq.~(\ref{ImSigma1}) if
$C_+(\alf) + C_-(\alf) = \frac{1}{\alf}$. In Ref.~\onlinecite{holder14} this latter
relation was reported only for the limit $\alf \gg 1$, while it is actually true for any $\alf$. Note that the momentum cutoff $\Lam$ in Eq.~(\ref{ImSigma1}) is not equivalent to the frequency cutoff $\Lam_\om$. Comparing Eqs.~(\ref{ImSigma1}) and (\ref{ImSigma2}) suggests that a momentum cutoff $\Lam$ corresponds to a frequency cutoff $\Lam_\om = \alf\Lam$. Anyway, numerical prefactors such as $\alf$ in the argument of the logarithm can also be absorbed in the subleading correction of order $k_0$.

The above results have been derived for the case where the Fermi surface is convex at the hot spots. For a concave Fermi surface, Eq.~(\ref{ImSigma1}) for $\Im\Sg(\bk_H,ik_0)$ remains the same, in Eq.~(\ref{ReSigma1}) for $\Re\Sg(\bk_H,ik_0) - \Sg(\bk_H,0)$ there is a sign change (from minus to plus) on the right hand side, and the expressions for $C_+(\alf)$ and $C_-(\alf)$ in Eq.~(\ref{C_alf}) are exchanged.

The logarithm in Eq.~(\ref{ImSigma1}) implies a logarithmic divergence in the one-loop contribution to the inverse quasi-particle weight,
$Z = 1 - \partial\Sg(\bk_H,ik_0)/(\partial ik_0) \sim
 \ln(\alf\Lam/|k_0|)$. \cite{footnote_Z}
Hence, Landau quasi-particles do not seem to exist at the hot spots.
Moreover, the self-energy exhibits an anomalous real part proportional to $|k_0|$. This contribution is directly related to the asymmetry of the real frequency self-energy with respect to a sign change of $\om$.


\subsection{Momentum dependence near hot spot}

We now discuss the momentum dependence of the one-loop self-energy near the hot spot $\bk_H$ at zero frequency, which we denote by $\Sg(k_r,k_t)$, where $k_r$ and $k_t$ are the normal and tangential momentum coordinates, respectively, relative to the hot spot, and the frequency argument has been dropped.

\begin{figure}
\centering
\includegraphics[width=8cm]{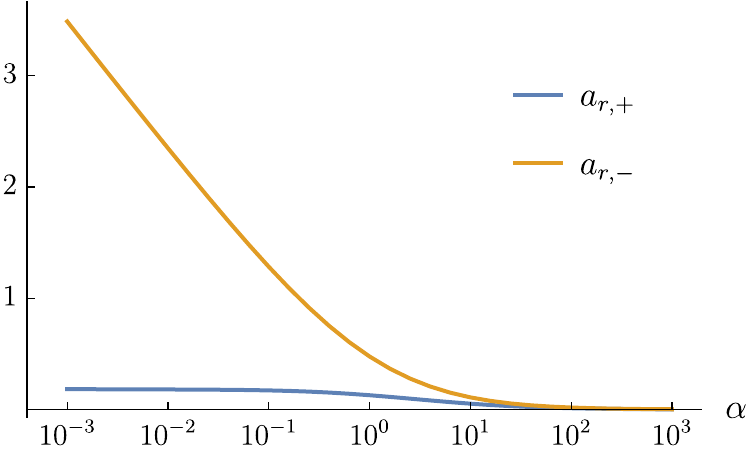}
\includegraphics[width=8cm]{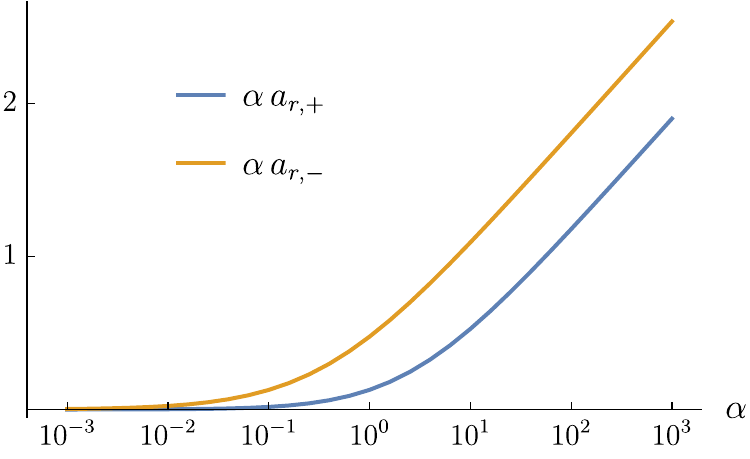}
\caption{The coefficients $a_{r,+}$ and $a_{r,-}$ as a function of the parameter $\alf$.
 On the right we plot $\alf a_{r,\pm}$ to show the behavior for large $\alf$.}
\label{fig:a_pm}
\end{figure}
The leading {\em normal}\/ momentum dependence for small $k_r$ has the form
\begin{equation} \label{Sigmak1}
 \Sg(k_r,0) - \Sg(0,0) = \frac{M}{N} \, a_{r,s_{k_r}}(\alf) \, v_F k_r \, ,
\end{equation}
where $a_{r,\pm}(\alf)$ are positive numbers depending on $\alf$ and on the sign of $k_r$. The derivation of this result and an expression for $a_{r,\pm}(\alf)$ in the form of a two-fold integral is provided in Appendix A. No ultraviolet cutoff is needed here.
In Fig.~\ref{fig:a_pm} we show the dependence of $a_{r,\pm}(\alf)$ on $\alf$. One can see that $a_{r,-}(\alf) > a_{r,+}(\alf)$ for all $\alf$.
For small $\alf$, one finds $a_{r,+} \sim \mbox{const}$ and $a_{r,-} \sim \log(1/\alf)$, while for large $\alf$ both coefficients $a_{r,\pm}$ are of order $\log(\alf)/\alf$. 
Due to the positive sign of $a_{r,\pm}(\alf)$, the self-energy enhances the bare Fermi velocity of the system. The enhancement is stronger for $k_r < 0$ than for $k_r > 0$, leading to a kink in the dispersion relation at the hot spot.
The above results have been derived for a convex Fermi surface. For a concave Fermi surface one obtains the same results with $a_{r,+}(\alf)$ and $a_{r,-}(\alf)$ exchanged.

The leading {\em tangential}\/ momentum dependence for small $k_t$ has the form
\begin{equation}
 \Sg(0,k_t) - \Sg(0,0) = \frac{M}{N} \, a_t(\alf) \sqrt{\Lam/m} \, k_t
 + {\cal O}(k_t^2) \, ,
\end{equation}
where $a_t(\alf)$ is a dimensionless function of $\alf$, and $\Lam$ is an ultraviolet cutoff on $e_\bq$ and $q_0$. The derivation of this result is provided in Appendix A.
The linear $k_t$ dependence of the self-energy entails a tilt of the Fermi surface at the hot spot. A deformation of the Fermi surface due to interactions is a generic phenomenon, not restricted to critical behavior. Since the hot spots and their antipodes on the Fermi surface are related by point group symmetries, the tilt does not spoil the $2k_F$ nesting condition of collinear Fermi velocities. Hence, the critical behavior is not affected by this tilt.

To analyze the contribution of order $k_t^2$, we have evaluated the second tangential momentum derivative of the self-energy $\partial_{k_t}^2 \Sigma(0,k_t)$ at $k_t = 0$. Applying this derivative to Eq.~(\ref{Sigma2}) and pulling it under the integral leads to an ill-defined expression due to multiple poles for zero frequency and momenta on the Fermi surface. The singularity can be regularized by introducing an infrared frequency cutoff $\Lam_{\rm IR}$. Qualitatively, this regularization corresponds to a finite temperature $T$, since the lowest fermionic Matsubara frequencies are given by $\pm \pi T$ for $T>0$. The resulting integral is convergent for any $k_t$, including $k_t = 0$, and it does not require any ultraviolet cutoff. A numerical evaluation of the integral shows that the regularized integral converges to a finite number even in the limit $\Lam_{\rm IR} \to 0$ (taken after integration). Hence, there is only a finite renormalization of the quadratic $k_t$ dependence of the fermionic dispersion. The mass $m$ and the Fermi surface curvature thus remain finite.


\section{Self-consistency check}

The self-energy obtained from the perturbative one-loop calculation modifies the fermion propagator substantially at the hot spots. Since the effective interaction $D$ and the fermion self-energy were computed with bare fermion propagators, we need to check whether the results remain qualitatively the same when the bare propagator is replaced by the interacting propagator with the non-Fermi liquid form
\begin{equation} \label{G}
 G(\bk,ik_0) = \frac{1}{Z(k_0) ik_0 + Z'|k_0| - \bar\xi_\bk}
\end{equation}
for low frequencies $k_0$ and momenta $\bk$ near a hot spot. From the Dyson equation $G^{-1} = G_0^{-1} - \Sg$ and Eqs.~(\ref{ImSigma1}) and (\ref{ReSigma1}) one can read off the renormalization factors
\begin{eqnarray}
 Z(k_0) &=& 1 + \frac{M}{\pi N \alf} \, \ln\frac{\alf\Lam}{|k_0|} \, , \label{Z} \\
 Z' &=& \frac{M}{\pi N \alf} \, R(\alf) \, . \label{Z'}
\end{eqnarray}
The dispersion relation $\bar\xi_\bk$ in Eq.~(\ref{G}) has a renormalized Fermi velocity following from the momentum dependence of the self-energy in Eq.~(\ref{Sigmak1}),
\begin{equation} \label{v_F}
 \bar v_{F,s_{k_r}} = v_F + \frac{M}{N} a_{r,s_{k_r}} v_F \, , 
\end{equation}
with a distinct renormalization for momenta inside and outside the Fermi surface, and a renormalized mass $\bar m$.

Eq.~(\ref{G}) describes the frequency dependence for $k_0 \to 0$ only for hot spot momenta. For momenta near a hot spot the frequency dependence is the same only above a momentum-dependent crossover scale, below which Fermi liquid behavior with a large finite $Z$ and a vanishing $Z'$ is recovered. The crossover scale is of order $k_r$ in the normal direction, and of order $k_t^2$ in the tangential direction. Approximating the fermion propagator by Eq.~(\ref{G}) in quantities that involve momentum integrals of $G(\bk,ik_0)$ will lead to quantitative errors, but it is not expected to affect the qualitative behavior.

In a self-consistent one-loop calculation, the effective interaction $D$ is still given by the RPA form
\begin{equation} \label{D_sc}
 D(\bq,iq_0) = \frac{g}{1 - g \Pi(\bq,iq_0)}
\end{equation}
as in Eq.~(\ref{D_rpa}), where the particle-hole bubble
\begin{equation} \label{Pi}
 \Pi(\bq,iq_0) = N \int \frac{d^2\bk}{(2\pi)^2} \int \frac{dk_0}{2\pi} \,
 G(\bk,ik_0) \, G(\bk-\bq,ik_0-iq_0) \, ,
\end{equation}
is now being computed with the interacting propagator $G$ instead of the bare one.
Vertex corrections do not play an important role here, since there are no singular one-loop vertex corrections for incommensurate density waves. \cite{altshuler95} For ordering wave vectors distinct from half a reciprocal lattice vector, there is no coalescence of divergences of propagators in the vertex correction.

To extract the leading momentum and frequency dependence of $\Pi(\bq,iq_0)$ for momenta near $\bQ$ and low frequencies, we expand the dispersion $\bar\xi_\bk$ to linear order in $k_r$ and to quadratic order in $k_t$ as previously. We first discard the (finite) renormalizations of the dispersion and replace $\bar\xi_\bk$ by the bare dispersion $\xi_\bk$, focusing thus on the changes imposed by the frequency dependence of the self-energy. The momentum integrals in Eq.~(\ref{Pi}) can then be computed analytically. In Appendix B we derive the result
\begin{eqnarray} \label{deltaPi}
 \delta\Pi(\bq,iq_0) &=& \Pi(\bq,iq_0) - \Pi(\bQ,0) \nonumber \\[1mm] &=&
 - N \frac{\sqrt{m}}{4\pi v_F} \left[
 \int dk_0 \, \frac{|\Theta(k_0 - q_0) - \Theta(-k_0)|}
 {\sqrt{-e_\bq + \{k_0\} + \{k_0 - q_0\}}} -
 \int \frac{dk_0}{\sqrt{2\{k_0\}}} \right] \, ,
\end{eqnarray}
where $\{k_0\} = Z(k_0)ik_0 + Z'|k_0|$.
For a numerical evaluation of the frequency integral in Eq.~(\ref{deltaPi}), we extend Eq.~(\ref{G}) to large frequencies by the ansatz
$Z(k_0) = 1 + \bar Z \ln\big(1 + \Lam/|k_0|\big)$ and $Z' = \bar Z'/(1 + |k_0|/\Lam)$,
where $\bar Z$ and $\bar Z'$ are constants, and $\Lam$ is an arbitrary ultraviolet cutoff. In this way the propagator has the correct large frequency asymptotics $G \sim (ik_0)^{-1}$ for $k_0 \to \infty$.
Eqs.~(\ref{Z}) and (\ref{Z'}) imply that the ratio $\bar Z'/\bar Z$ is given by $R(\alf)$.

\begin{figure}
\centering
\includegraphics[width=10cm]{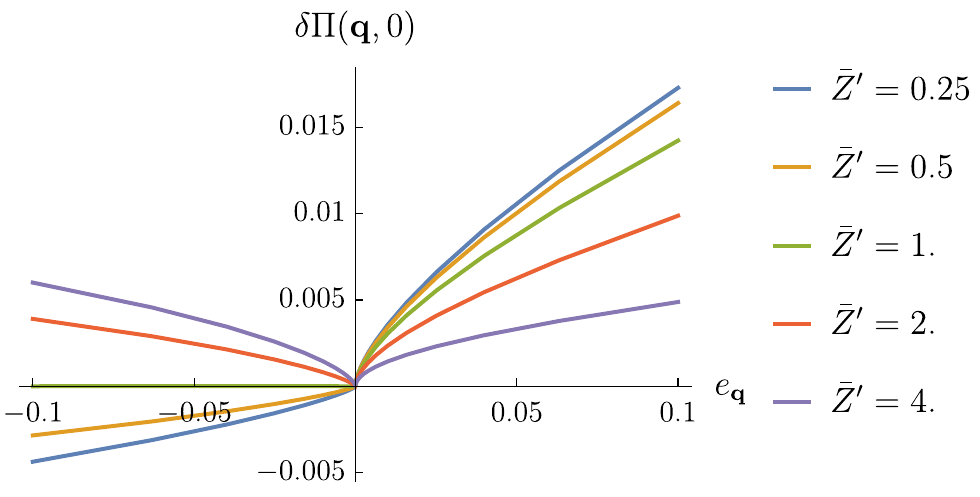}
\caption{Singular part of the static particle-hole bubble $\Pi(\bq,0)$ as a function
 of $e_\bq$ as obtained from fermion propagators with $Z$-factors resulting
 from the frequency dependence of the one-loop self-energy.
 The various curves correspond to distinct choices of $\bar Z'$,
 while $\bar Z = 2/\pi$ is fixed. The UV cutoff $\Lam$ has been set to one.
 The other parameters are $N = m = v_F = 1$.}
\label{fig:Pi_pert}
\end{figure}
In Fig.~\ref{fig:Pi_pert} we plot the static ($q_0=0$) particle-hole bubble $\delta\Pi(\bq,0)$ as a function of $e_\bq$ for $\bar Z = 2/\pi$ and various choices of $\bar Z'$. One can see that $\delta\Pi(\bq,0)$ exhibits a peak at $\bq = \bQ$ only if $\bar Z' > 1$. Since the behavior for small $e_\bq$ is determined by low frequency contributions, the criterion for a peak is independent of the choice of $\Lam$ and depends only on the ratio $\bar Z'/\bar Z$. Hence, a peak is obtained if and only if this ratio is larger than $\pi/2$. However, the perturbative result yields $\bar Z'/\bar Z = R(\alf)$ which is smaller than one. The one-loop fermion self-energy computed with bare propagators thus removes the peak in the particle-hole bubble at the nesting vector, and thus the corresponding peak in the susceptibility and in the effective interaction $D$. The $2k_F$ quantum critical point seems thus spoiled by the self-energy feedback, and the perturbative calculation is not even qualitatively consistent.

We now analyze the effect of the Fermi velocity renormalization (\ref{v_F}) generated by the momentum dependence of the self-energy. Although this is a finite renormalization, it leads to a qualitative change since the renormalization factors differ for particles inside and outside the Fermi sea. We compute $\delta\Pi(\bq,iq_0)$ for momenta near $\bQ$ and low frequencies as previously, but now with the renormalized Fermi velocity as obtained from the one-loop self-energy. We still ignore the finite renormalization of the mass. In this case only the $k_t$ integration in Eq.~(\ref{Pi}) can be performed analytically by residues, but the remaining two integrals can be easily carried out numerically.

\begin{figure}
\centering
\includegraphics[width=8cm]{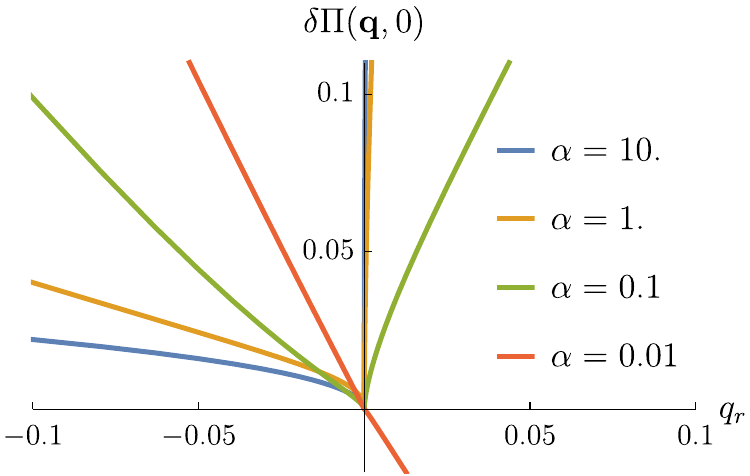}
\caption{Singular part of the static particle-hole bubble $\Pi(\bq,0)$ as a function
 of $q_r$ at $q_t=0$ as obtained from fermion propagators with renormalizations resulting
 from the momentum and frequency dependence of the one-loop self-energy.
 The various curves correspond to distinct choices of the parameter $\alf$.
 The other parameters are $M = N = m = v_F = 1$.}
\label{fig:Pi_pert'}
\end{figure}
In Fig.~\ref{fig:Pi_pert'} we show the resulting static particle-hole bubble $\delta\Pi(\bq,0)$ as a function of $q_r$ for various choices of the parameter $\alf$. The $Z$-factors in the fermion propagators assume the $\alf$-dependent values described by Eqs.~(\ref{Z}) and (\ref{Z'}) at low frequencies. The renormalized Fermi velocity is given by Eq.~(\ref{v_F}) for all momenta. One can see that that $\Pi(\bq,0)$ now exhibits a sharp peak at $\bQ$ for sufficiently large values of $\alf$. The peak at $\bQ$ is again destroyed for small $\alf$, but this time due to a change of slope for positive $q_r$. The critical value for the stability of the peak at $\bQ$ is $\alf_c \approx 0.03$. Note that this value depends on the choice of $M$ and $N$. For $M=1$ and $N=2$ we find $\alf_c \approx 1$.
In any case, the Fermi velocity renormalization with a larger renormalization factor for particles inside the Fermi sea helps in stabilizing the QCP with a nested wave vector. Indeed, taking only the Fermi velocity renormalization into account, and ignoring the frequency dependence of the self-energy, we found that the peak at $\bQ$ is stabilized for any choice of $\alf$.

While the QCP seems stabilized at least for sufficiently large $\alf$, the one-loop self-energy feedback leads to a steeper, non-linear momentum dependence of $\Pi(\bq,0)$ for negative $q_r$. This, in turn, reduces the singular contributions to the self-energy, so that the perturbative one-loop result with bare propagators is not self-consistent anyway.

In the following two sections we try to incorporate the self-energy feedback self-consistently, first by a renormalization group flow, and then by a self-consistent solution of the coupled integral equations for the particle-hole bubble and the fermion self-energy. In both sections we focus on the frequency dependence of the self-energy, discarding its momentum dependence for simplicity.


\section{Renormalization group analysis}

A systematic way of dealing with low energy singularities and the corresponding divergences in perturbation theory is provided by the renormalization group, where self-consistency can in principle be achieved by solving a set of flow equations, that is, ordinary differential equations, instead of solving non-linear integral equations.
In this section we derive and solve a flow equation for the fermion self-energy with self-energy feedback on the right hand side, and we compare the results to those from perturbation theory. To access the full frequency dependence of the self-energy, we use a {\em functional}\/ renormalization group (fRG) framework. \cite{metzner12}
Flow equations based on the fRG have already been applied to other quantum critical points in two-dimensional interacting fermion systems, for example, at the onset of nematic \cite{drukier12} and (non-nested) antiferromagnetic \cite{maier16} order in metals, and at the onset of superfluidity in semi-metals. \cite{obert11}


\subsection{Flow equation}

To capture order parameter fluctuation effects efficiently, it is convenient to decouple the two-fermion interaction by introducing a bosonic order parameter field via a Hubbard-Stratonovich transformation. \cite{stratonovich58,hubbard59} The system is then described by a coupled fermion-boson theory with a bare fermion propagator $G_0(\bk,ik_0) = [ik_0 - \xi_\bk]^{-1}$, a bare boson progagator $D_0(\bq,iq_0) = g$, and a constant Yukawa interaction which couples the fermionic charge or spin density linearly to the boson field. The bare Yukawa coupling is equal to one.

The fRG is based on a scale-by-scale evaluation of the functional integral representing the partition function and correlation functions of the system. \cite{metzner12} A flow is generated by letting the bare propagator depend on a flow parameter $\Lam$, usually an infrared cutoff. The corresponding scale-dependent effective action $\Gamma^\Lam$ interpolates smoothly between the bare action of the system and the final effective action $\Gamma = \lim_{\Lam \to 0} \Gamma^\Lam$, from which the grand canonical potential, the self-energy, and higher order vertex functions can be obtained. The flow of $\Gamma^\Lam$ is governed by an exact functional flow equation. \cite{wetterich93}

We impose a sharp frequency cutoff on the fermion propagator, that is,
\begin{equation}
 G_0^\Lam(\bk,ik_0) = \frac{\Theta(|k_0| - \Lam)}{ik_0 - \xi_\bk} \, .
\end{equation}
This suppresses all contributions to the functional integral at the initial cutoff $\Lam_0 = \infty$, and it regularizes the Fermi surface singularity at $k_0 = \xi_\bk = 0$ until $\Lam \to 0$. No cutoff is needed for the bare boson propagator $D_0$, since it is bounded, and also the full boson propagator remains finite as long as $\Lam > 0$.

The exact functional flow equation for $\Gamma^\Lam$ leads to an infinite hierarchy of flow equations for the self-energies (for fermions and bosons) and vertex functions of any order. \cite{metzner12} We truncate this hierarchy at the leading order, that is, we keep only the self-energies and the Yukawa vertex.
For incommensurate density waves, the Yukawa vertex receives no singular contributions because of the absence of coalescent divergences in the vertex correction. \cite{altshuler95} Neglecting its flow altogether does therefore not affect the qualitative behavior of the self-energies. We thus fix the Yukawa coupling at its initial value (one).

The scale-dependent self-energies are related to the full and bare propagators by the usual Dyson equations, that is
$G^\Lam = \left[(G_0^\Lam)^{-1} - \Sg^\Lam \right]^{-1}$ and
$D^\Lam = \left[ (D_0)^{-1} - \Pi^\Lam \right]^{-1}$, where $\Sg^\Lam$ and $\Pi^\Lam$ are the fermionic and bosonic self-energies, respectively.
The flow equation for the fermion self-energy reads
\begin{equation} \label{floweq_Sg}
 \partial_\Lam \Sg^\Lam(\bk,ik_0) = 
 - M \int \! \frac{d^2\bq}{(2\pi)^2} \int \! \frac{dq_0}{2\pi} \,
 D^\Lam(\bq,iq_0) \, S^\Lam(\bk-\bq,ik_0-iq_0) ,
\end{equation}
where $S^\Lam$ is the single-scale propagator \cite{metzner12}
\begin{equation} \label{S_Lam}
 S^\Lam(\bk,ik_0) =
 \left. \frac{\partial G^\Lam(\bk,ik_0)}{\partial\Lam}
 \right|_{\Sg^\Lam = {\rm const}} =
 - \frac{\delta(|k_0| - \Lam)}{ik_0 - \xi_\bk - \Sg^\Lam(\bk,ik_0)} .
\end{equation}
The flow equation for the boson self-energy has the form
\begin{eqnarray} \label{floweq_Pi}
 \partial_\Lam \Pi^\Lam(\bq,iq_0) &=&
 N \int \! \frac{d^2\bk}{(2\pi)^2} \int \! \frac{dk_0}{2\pi} \big[ 
 S^\Lam(\bk,ik_0) \, G^\Lam(\bk-\bq,ik_0-iq_0) \nonumber \\[2mm]
 && + \, G^\Lam(\bk,ik_0) \, S^\Lam(\bk-\bq,ik_0-iq_0) \big] .
\end{eqnarray}
These flow equations can be derived by inserting the truncation described above into the exact hierarchy of flow equations. \cite{metzner12} Note that they have the same form as a $\Lam$-derivative of the perturbative expressions Eqs.~(\ref{Sigma1}) and (\ref{Pi}), with full and scale-dependent propagators, and the proviso that on the right hand sides the derivative acts only on the explicit cutoff dependence of $G^\Lam$. 

Due to the $\delta$-function in the single-scale propagator $S^\Lam$, the frequency integral in Eq.~(\ref{floweq_Sg}) can be easily carried out, leading to
\begin{equation}
 \partial_\Lam \Sg^\Lam(\bk,ik_0) =
 \frac{M}{2\pi} \sum_{s = \pm 1} \int \frac{d^2\bq}{(2\pi)^2} \,
 \frac{D^\Lam(\bq,ik_0 - is\Lam)}{is\Lam - \xi_{\bk-\bq} - \Sg^\Lam(\bk-\bq,is\Lam)}
 \, .
\end{equation}
We focus on the frequency dependence of the self-energy at a hot spot as described by the function $\Sg_H^\Lam(ik_0) = \Sg^\Lam(\bk_H,ik_0) - \Sg^\Lam(\bk_H,0)$. Neglecting the $\bq$-dependence of the self-energy on the right hand side of the flow equation, and absorbing the constant $\Sg^\Lam(\bk_H,0)$ by a shift of the chemical potential (keeping the Fermi surface fixed), we obtain the simplified flow equation
\begin{equation} \label{floweq_Sg_H}
 \partial_\Lam \Sg_H^\Lam(ik_0) =
 \frac{M}{2\pi} \sum_{s = \pm 1} \int \frac{d^2\bq}{(2\pi)^2} \,
 \frac{D^\Lam(\bq,ik_0 - is\Lam) - D^\Lam(\bq,-is\Lam)}
 {is\Lam - \xi_{\bk_H-\bq} - \Sg_H^\Lam(is\Lam)}
 \, .
\end{equation}
The singular contributions to the fermion self-energy are due to momentum transfers $\bq$ near $\bQ$. We therefore can use the coordinates $q_r$ and $q_t$ as in the perturbative calculation, and expand $\xi_{\bk_H-\bq} = v_F q_r + \frac{1}{2m} q_t^2$. An ultraviolet cutoff $\Lam_\bq$ restricts the momentum integral to $|q_r| \leq \Lam_\bq$ and $|q_t| \leq \Lam_\bq$.

To compute the boson self-energy $\Pi^\Lam$, we partition the $\bk$-integration domain in Eq.~(\ref{floweq_Pi}) in regions close to the hot spots $\bk_H$ and $\bk'_H$ (so that $\bk-\bq$ is close to $-\bk_H$ and $-\bk'_H$, respectively), and regions far from the hot spots. The singular part of $\Pi^\Lam$ is entirely due to the former regions. We denote the contribution from $\bk$ near $\bk_H$ as $\Pi_H^\Lam$ and the contribution from $\bk$ near $\bk'_H$ as $\Pi_{H'}^\Lam$. The total self-energy can thus be written as $\Pi^\Lam = \Pi_H^\Lam + \Pi_{H'}^\Lam + \Pi_{\rm reg}^\Lam$, where the last contribution comes from momenta $\bk$ far from $\bk_H$ and $\bk'_H$ and is regular even for $\Lam \to 0$.
For $\bq$ near $\bQ$, the regular part can be expanded to linear order in $e_\bq$ and $q_t$ as $\Pi_{\rm reg}^\Lam(\bq,iq_0) = \Pi_{\rm reg}^\Lam(\bQ,0) - N(b^\Lam e_\bq + c^\Lam q_t$). Its frequency dependence is irrelevant compared to the singular terms.

The flow equations for $\Pi_H^\Lam$ and $\Pi_{H'}^\Lam$ containing the singular contributions can be simplified by expanding the dispersion $\xi_\bk$ around the hot spots. For example, for $\bk$ near $\bk_H$, we expand
$\xi_\bk = v_F k_r + \frac{1}{2m} k_t^2$ and
$\xi_{\bk-\bq} = - v_F (k_r-q_r) + \frac{1}{2m} (k_t-q_t)^2$. Inserting this expansion, the momentum integral in the flow equation (\ref{floweq_Pi}) is convergent without the need of an ultraviolet cutoff. For a momentum-independent self-energy, the momentum integral can be performed analytically. Extending the integration region of $k_r$ and $k_t$ to infinity does not affect the low frequency behavior, and allows for an easy evaluation via the residue theorem.
The calculation is basically the same as the one leading to Eq.~(\ref{PiA1}) in Appendix B, and yields
\begin{eqnarray} \label{floweq_Pi_H}
 \partial_\Lam \Pi_H^\Lam(\bq,iq_0) &=& N \frac{\sqrt{m}}{4\pi v_F} \int dk_0 \,
 \frac{|\Theta(k_0-q_0) - \Theta(-k_0)|}{\sqrt{-e_\bq + \{k_0\} + \{k_0-q_0\}}}
 \nonumber \\[2mm]
 && \times \big[ \delta(|k_0|-\Lam) \Theta(|k_0-q_0|-\Lam) +
 \Theta(|k_0|-\Lam) \delta(|k_0-q_0|-\Lam) \big]
 \nonumber \\[2mm]
 &=& N \frac{\sqrt{m}}{4\pi v_F} \sum_{s = \pm 1}
 \frac{1}{\sqrt{-e_\bq + \{s\Lam\} + \{s(\Lam + |q_0|)\}}} \, ,
\end{eqnarray}
where $\{k_0\} = ik_0 - \Sg_H^\Lam(k_0)$. The flow equation for $\Pi_{H'}^\Lam(\bq,iq_0)$ has the same form with $e_\bq$ replaced by $e'_\bq$.

For large $\Lam$, the flow generates only smooth contributions. Hence, we do not start the flow at $\Lam_0 = \infty$, but rather at some finite $\Lam_0$. In the regular part of $\Pi^\Lam$ we discard the contributions from $\Lam < \Lam_0$ and insert fixed (scale-independent) parameters for $\Pi_{\rm reg}^\Lam(\bQ,0)$, $b^\Lam$, and $c^\Lam$. By this procedure we just miss some finite renormalizations. 
Hence, the scale-dependent boson propagator has the final form
\begin{equation} \label{D_Lam}
 D^\Lam(\bq,iq_0) =
 \left[ \bar g^{-1} - \Pi_{H}^\Lam(\bq,iq_0) - \Pi_{H'}^\Lam(\bq,iq_0)
 + N (b \, e_\bq + c \, q_t) \right]^{-1} \, ,
\end{equation}
where $\bar g^{-1} = g^{-1} - \Pi_{\rm reg}(\bQ,0)$, and the flow of $\Pi_{H}^\Lam(\bq,iq_0)$ and $\Pi_{H'}^\Lam(\bq,iq_0)$ is determined by Eq.~(\ref{floweq_Pi_H}).
To tune to the quantum critical point, the renormalized coupling $\bar g$ has to be chosen such that 
\begin{equation} \label{g_bar}
 \bar g^{-1} = \lim_{\Lam \to 0} \left[ \Pi_{H}^\Lam(\bQ,0) + \Pi_{H'}^\Lam(\bQ,0)
 \right] .
\end{equation}
%


\subsection{Results}

We now show results for the fermion self-energy and the boson propagator as obtained from a numerical solution of the flow equations. We choose a fixed set of parameters $M = N = v_F = v'_F = m = m' = 1$, and $b = c = 1/(8\pi)$. The angle $\theta$ between the Fermi surface normal vectors at $\bk_H$ and $\bk'_H$ is chosen as $\pi/2$, such that $q'_r = q_t$ and $q'_t = -q_r$. \cite{footnote_num}
For the initial value of the frequency cutoff we choose $\Lam_0 = 10$, and the momentum integrals for the fermion self-energy are cut off by $\Lam_\bq = 1$. Other choices of the cutoffs lead to the same qualitative behavior.

The coupling constant $\bar g$ is tuned to the quantum critical point, that is, we try to choose it such that Eq.~(\ref{g_bar}) is satisfied at the end of the flow. However, it turns out that the effective interaction diverges at a wave vector $\bar\bQ$ slightly away from the nesting vector $\bQ$ (see below). Hence, we show results where $\lim_{\Lam \to 0} \big[ D^\Lam(\bar\bQ,0)]^{-1} = 0$. The corresponding coupling constant is $\bar g_c = - 1.00535$.

\begin{figure}
\centering
\includegraphics[width=9cm]{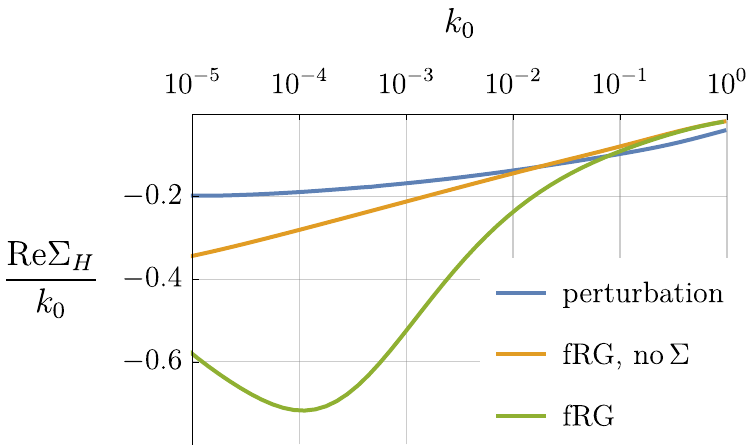} \\
\includegraphics[width=9cm]{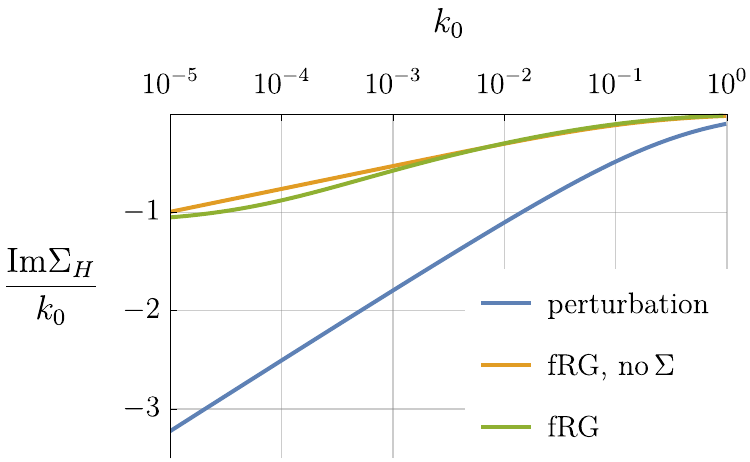}
\caption{Real (top) and imaginary (bottom) parts of the fermion self-energy at
 a hot spot, divided by $k_0$, as a function of the Matsubara frequency $k_0$.
 The result from the flow equations is compared to the result from simplified
 flow equations without self-energy feedback, and to the result from one-loop
 perturbation theory.}
\label{fig:Sg_fRG}
\end{figure}
In Fig.~\ref{fig:Sg_fRG} we show the result for the fermion self-energy at a hot spot as a function of frequency. We plot the ratio $\Sg_H(k_0)/k_0$ to reveal the deviations from a linear frequency dependence. We also show the result obtained from flow equations with bare fermion propagators, that is, without fermion self-energy feedback (here the ordering wave vector remains $\bQ$), as well as the result from plain one-loop perturbation theory. The latter can be obtained from the flow equation for the fermion self-energy by neglecting the self-energy feedback and by inserting the fully integrated bare particle-hole bubble into the flow equation for the fermion self-energy.

The perturbative result agrees with the asymptotic low frequency analysis presented in Sec.~III. The real part of $\Sg_H(k_0)$ is linear in frequency at low frequencies, and the imaginary part is proportional to $k_0 \log k_0$, in agreement with Eqs.~(\ref{ImSigma1}) and (\ref{ReSigma1}). For $k_0 \to 0$ the ratio $\Re\Sg_H(k_0)/k_0$ tends to a finite value near $-0.2$ in agreement with the asymptotic result $-R(\alf)/(\pi\alf) = -0.209$ for $\alf = 8\pi v_F c = 1$. The asymptotic behavior of the real part is approached rather slowly, because subleading corrections to the leading low frequency behavior are suppressed only by a low power of frequency.

The self-energy obtained from the flow equation (with self-energy feedback) does not exhibit any simple scaling behavior, neither at intermediate nor at the lowest accessible frequencies. The imaginary part is only roughly proportional to $k_0 \log k_0$, where the prefactor is significantly smaller than in perturbation theory, and it starts decreasing for frequencies below $k_0 \approx 10^{-4}$ instead of tending to a constant.
$\Re\Sg_H(k_0)/k_0$ exhibits a pronounced frequency dependence both at high and low frequencies. There is a minimum at $k_0 \approx 10^{-4}$.
We have checked that the change of trends in both real and imaginary parts at $k_0 \approx 10^{-4}$ is not an artifact of the final infrared cutoff $\Lambda_f = 10^{-8}$ at which we have to stop the flow before running into numerical instabilities.

Discarding the self-energy feedback on the right hand side of the flow equations has a significant effect on the frequency dependence. Both real and imaginary parts of the self-energy obtained from such a simplified flow are proportional to $k_0 \log k_0$ over a wide frequency range down to the lowest accessible frequencies. Hence, the change of trend at $k_0 \approx 10^{-4}$ observed above is obviously due to the self-energy feedback.

\begin{figure}
\centering
\includegraphics[width=8.5cm]{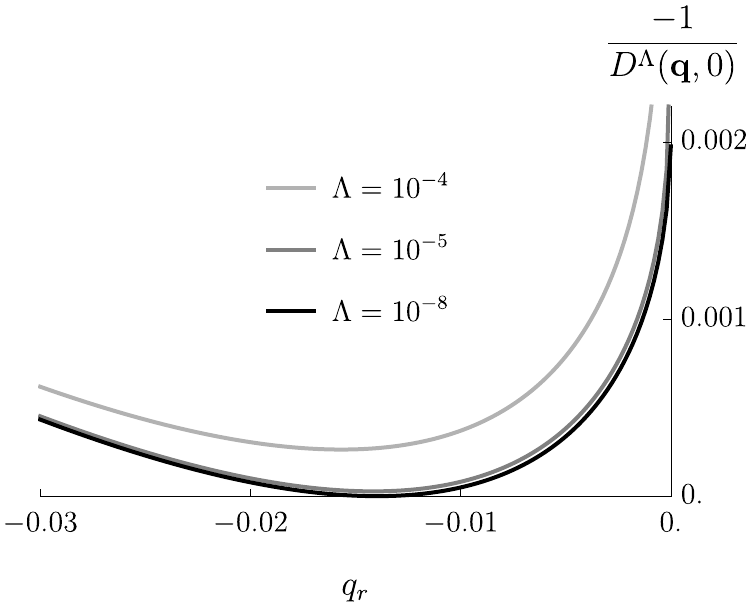}
\caption{Inverse fluctuation propagator $[D^{\Lam}(\bq,0)]^{-1}$ at three stages
 of the flow for $q_t = q_r$ as a function of $q_r$.
 The data have been obtained from the flow equations with self-energy feedback.}
\label{fig:D_fRG}
\end{figure}
In Fig.~\ref{fig:D_fRG} we show the momentum dependence of the inverse fluctuation propagator $[D^{\Lam}(\bq,0)]^{-1}$ at three stages of the flow ($\Lam = 10^{-4}$, $10^{-5}$, and $\Lam_f = 10^{-8}$) for $q_t = q_r$ as a function of $q_r$. One can see that $[D^{\Lam_f}(\bq,0)]^{-1}$ vanishes at a small negative value of $q_r = q_t \approx - 0.014$, while it is negative everywhere else. It is negative also for momenta with $q_r \neq q_t$ (not shown in the figure).
The fluctuation propagator thus diverges at a wave vector $\bar\bQ \neq \bQ$. The $2k_F$ QCP is thus spoiled, and the low energy behavior will ultimately be governed by a different universality class with a non-nested ordering wave vector.
Since the shift of the ordering wave vector is very small, its effect on the fermion self-energy will be visible only at very low frequencies. We do not explore this in more detail, since there are reasons to believe that the shift of $\bQ$ is an artifact of the approximate flow equations, as we will now explain.

\begin{figure}
\centering
\includegraphics[width=10cm]{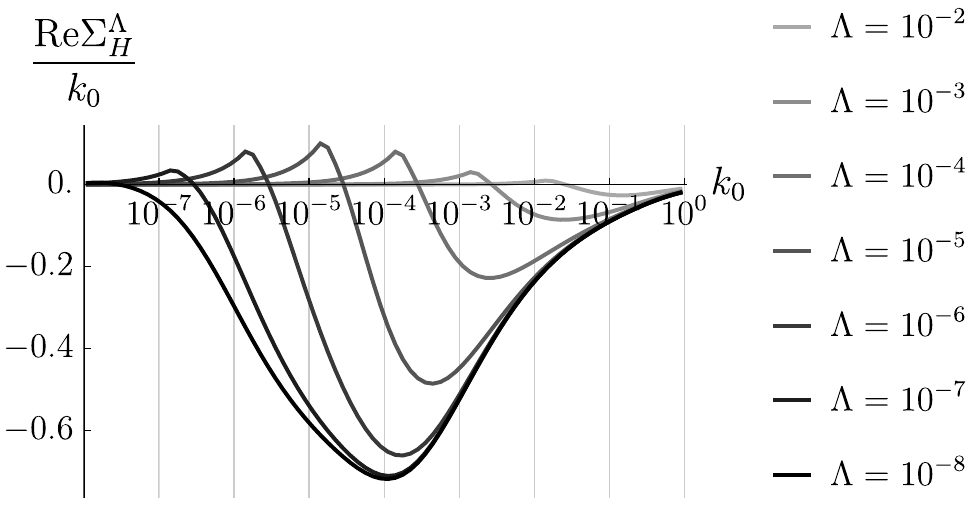} \\[5mm]
\includegraphics[width=10cm]{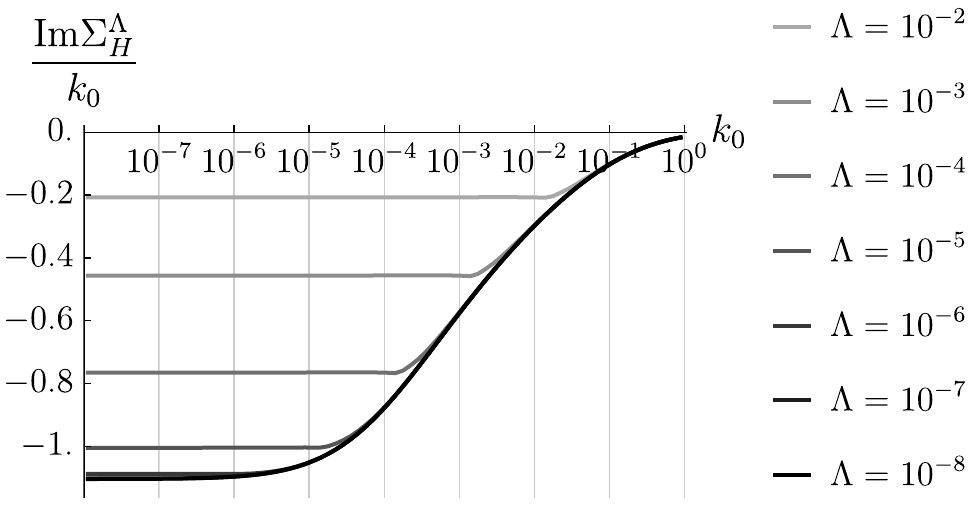}
\caption{Real (top) and imaginary (bottom) parts of the fermion self-energy at
 a hot spot, divided by $k_0$, as a function of the Matsubara frequency $k_0$.
 The results from the flow equation (with self-energy feedback) are shown at
 various stages of the flow.}
\label{fig:Sg_Lam}
\end{figure}
The real part of the self-energy exhibits a very peculiar cutoff dependence. In Fig.~\ref{fig:Sg_Lam} we show how the self-energy evolves in the course of the flow. $\Im\Sg_H^\Lam(ik_0)$ converges to its final value (for $\Lam \to 0$) at a scale $\Lam$ of order $k_0$. By contrast, $\Re\Sg_H^\Lam(ik_0)$ converges much slower, roughly at scales $\Lam \sim 10^{-3} k_0$. For $\Lam \sim k_0$ the real part of the self-energy is very far from $\lim_{\Lam \to 0} \Re\Sg_H^\Lam(ik_0)$, it even has the opposite sign.

On the right hand side of the flow equations the self-energy is evaluated for frequencies at or near $\Lam$. Hence, the real part of the self-energy inserted on the right hand side of the flow equations differs drastically from the real part of the final self-energy (for $\Lam \to 0$), it even has the wrong sign. This is a very unusual situation, and it seems that such a self-energy feedback works against self-consistency rather than implementing it. Choosing a smooth instead of a sharp cutoff doesn't help. At present we do not know how to improve the truncation of the exact fRG flow equation to avoid this problem.


\section{Self-consistent solution}

We now present a self-consistent solution of the coupled one-loop equations for the fermion and boson self-energies. Momentum dependences will be approximated in a similar spirit as in the previous section. We will also introduce an infrared frequency cutoff $\Lam$ as in the fRG approach. Here, this cutoff is not used for setting up differential flow equations. It is merely a technical device to achieve a self-consistent solution by iteration. 
We will see that the $2k_F$ nesting QCP persists in the self-consistent solution.


\subsection{Self-consistent equations}

The self-consistent one-loop equations are given by
\begin{equation} \label{Sigma_sc}
 \Sg(\bk,ik_0) = - M \int \frac{d^2\bq}{(2\pi)^2} \int \frac{dq_0}{2\pi} \,
 D(\bq,iq_0) \, G(\bk-\bq,ik_0-iq_0) \, ,
\end{equation}
where $G(\bk,ik_0)$ is the full propagator, and Eq.~(\ref{D_sc}) for the effective interaction $D(\bq,iq_0)$ with the particle-hole bubble $\Pi(\bq,iq_0)$ from Eq.~(\ref{Pi}). Together these equations form a coupled set of non-linear integral equations which are very hard to solve with the accuracy needed to resolve the low frequency behavior. We therefore focus on the momentum region near the hot spots, and we simplify the equations by expanding the momentum dependences in close analogy to what we did in the previous sections.

We define $\Sg_H(ik_0) = \Sg(\bk_H,ik_0) - \Sg(\bk_H,0)$, neglect the $\bq$-dependence of the self-energy on the right hand side of Eq.~(\ref{Sigma_sc}), and absorb the constant $\Sg(\bk_H,0)$ by a shift of the chemical potential. Introducing a sharp infrared frequency cutoff $\Lam$, we then obtain
\begin{equation} \label{sceq_Sg_H}
 \Sg_H^\Lam(ik_0) =
 - M \int \frac{dk'_0}{2\pi} \, \Theta_\Lam^{\Lam_0}(k'_0)
 \int \frac{d^2\bq}{(2\pi)^2} \,
 \frac{D^\Lam(\bq,i(k_0 - k'_0)) - D^\Lam(\bq,-ik'_0)}
 {ik'_0 - \xi_{\bk_H-\bq} - \Sg_H^\Lam(ik'_0)} \, ,
\end{equation}
where $\Theta_\Lam^{\Lam_0}(k_0) = 1$ for $\Lam \leq |k_0| \leq \Lam_0$ and zero else. An ultraviolet momentum cutoff $\Lam_\bq$ restricts the momentum integration to $|q_r| \leq \Lam_\bq$ and $|q_t| \leq \Lam_\bq$. The dispersion is expanded as previously around the hot spot, that is, $\xi_{\bk_H - \bq} = v_F q_r + \frac{1}{2m} q_t^2$.

The particle-hole bubble is again decomposed as $\Pi = \Pi_H + \Pi_{H'} + \Pi_{\rm reg}$, where the first two contributions come from momenta close to $\bk_H$ and $\bk'_H$, respectively, and the last (regular) contribution from momenta far from $\bk_H$ and $\bk'_H$.
For $\bq$ near $\bQ$, the regular part can be expanded to linear order in $e_\bq$ and $q_t$ as $\Pi_{\rm reg}(\bq,iq_0) = \Pi_{\rm reg}(\bQ,0) - N(b e_\bq + c q_t$). Its frequency dependence is irrelevant compared to the singular terms.

The singular contributions $\Pi_H$ and $\Pi_{H'}$ are computed by expanding the dispersion around the hot spots and extending the momentum integrals to infinity. Eq.~(\ref{PiA1}) yields
\begin{equation} \label{sceq_Pi_H}
 \Pi_H^\Lam(\bq,iq_0) = - N \frac{\sqrt{m}}{v_F}
 \int_\Lam^{\Lam_0} \frac{dk_0}{2\pi} \, \Re \frac{1}
 {\sqrt{-e_\bq + 2ik_0 + i|q_0| - \Sg_H^\Lam(ik_0) - \Sg_H^\Lam(ik_0 + i|q_0|)}}
 \, .
\end{equation}
$\Pi_{H'}^\Lam(\bq,iq_0)$ has the same form with $e_\bq$ replaced by $e'_\bq$.
The boson propagator $D^\Lam(\bq,iq_0)$ is then given by Eq.~(\ref{D_Lam}) as in the fRG approach.

It is instructive to compare the integral equations (\ref{sceq_Sg_H}) and (\ref{sceq_Pi_H}) to the flow equations (\ref{floweq_Sg_H}) and (\ref{floweq_Pi_H}), respectively. The latter could be obtained from the former by applying a $\Lam$-derivative which acts only on the integration boundary imposed by $\Lam$ on the right hand side, not on the $\Lam$-dependent integrands.

We solve the coupled integral equations by reducing $\Lam$ gradually from $\Lam_0$ to the smallest accessible values, and updating the fermion self-energy at each step. Due to the gradual adaption of the self-energy in this procedure a converged solution can be achieved, while by a direct iteration without infrared cutoff it is difficult to reach the self-consistent attractor. The frequency cutoff $\Lam$ is thus a useful device to obtain a self-consistent solution. Ideally one would like to reduce it to zero at the end of the calculation. We managed to reduce it to $10^{-7.1}$ before running into numerical instabilities.


\subsection{Results}

We now show results for the fermion self-energy and the fluctuation propagator as obtained from a numerical solution of the self-consistent equations. We choose the same fixed set of parameters as in the fRG calculation in the previous section, that is, $M = N = v_F = m = 1$, $\theta = \pi/2$, and $b = c = 1/(8\pi)$. \cite{footnote_num} The ultraviolet cutoffs are also the same: $\Lam_0 = 10$ and $\Lam_\bq = 1$. The coupling constant $\bar g$ is tuned to the QCP, that is, it is chosen such that $D^\Lam(\bQ,0)$ diverges for $\Lam \to 0$, but not earlier. The critical coupling is $\bar g_c = - 1.00338$, which differs only slightly from the critical coupling obtained from the fRG.

\begin{figure}
\centering
\includegraphics[width=9cm]{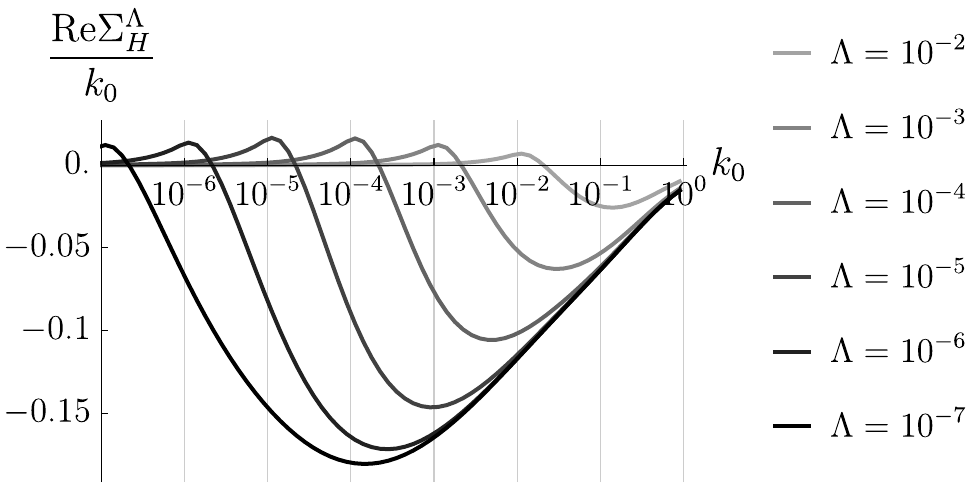} \\[5mm]
\includegraphics[width=9cm]{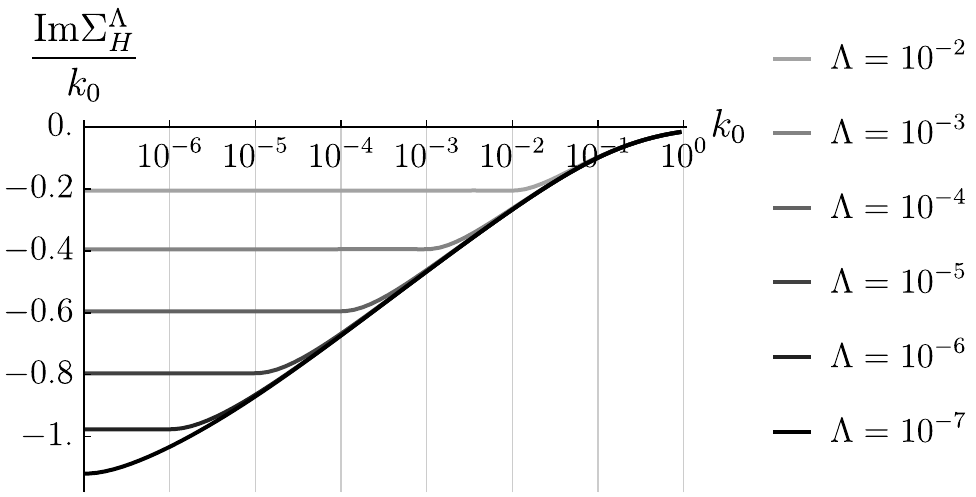}
\caption{Real (top) and imaginary (bottom) parts of the fermion self-energy at
 a hot spot, divided by $k_0$, as a function of the Matsubara frequency $k_0$.
 The various curves represent self-consistent solutions for various choices of
 the infrared cutoff $\Lam$.}
\label{fig:Sg_sc}
\end{figure}
In Fig.~\ref{fig:Sg_sc} we show the self-consistent result for the fermion self-energy at a hot spot as a function of frequency, plotting the ratio $\Sg_H(k_0)/k_0$, for various choices of the infrared cutoff $\Lam$. At first sight the results look similar to the fRG results (see Fig.~\ref{fig:Sg_Lam}), except for the overall size of the real part, which is smaller by a factor of four compared to the self-energy obtained from the flow equations. The convergence of $\Sg_H^\Lam(k_0)$ for fixed $k_0$ and $\Lam \to 0$ is again very slow for the real part. However, in contrast to the fRG flow, the self-energy feedback now involves all frequencies, not only those near the scale $\Lam$. Hence, in the limit $\Lam \to 0$, the cutoff dependence of the feedback disappears.

The converged imaginary part $\Im\Sg_H^{\Lam \to 0}(k_0)$ is proportional to $k_0 \log k_0$ in a broad frequency range $k_0 < 10^{-2}$. It is not clear whether the decreasing slope at the lowest frequencies (below $k_0 \approx 10^{-5}$) is exclusively due to the finite cutoff. The real part $\Re\Sg_H^{\Lam \to 0}(k_0)$ also seems proportional to $k_0 \log k_0$ in a certain frequency range, which is however limited to frequencies above $10^{-3}$. For each $\Lam$, $\Re\Sg_H^\Lam(k_0)/k_0$ exhibits a minimum at a frequency $k_0^*(\Lam)$ much larger than $\Lam$. The position of the minimum decreases with decreasing $\Lam$, but the ratio $k_0^*(\Lam)/\Lam$ increases. Hence, it is not clear whether $k_0^*(\Lam)$ approaches zero or a finite value for $\Lam \to 0$. In other words, we do not know whether the minimum of $\Re\Sg_H^\Lam(k_0)/k_0$ persists for $\Lam \to 0$. In view of the slow convergence of the real part for $\Lam \to 0$, and our limitation to cutoffs $\Lam > 10^{-7}$, we believe that our results for the real part are converged only for frequencies $k_0 \geq 10^{-3}$, while for the imaginary part the results seem converged for $k_0 \geq 10^{-5}$.

\begin{figure}
\centering
\includegraphics[width=8.5cm]{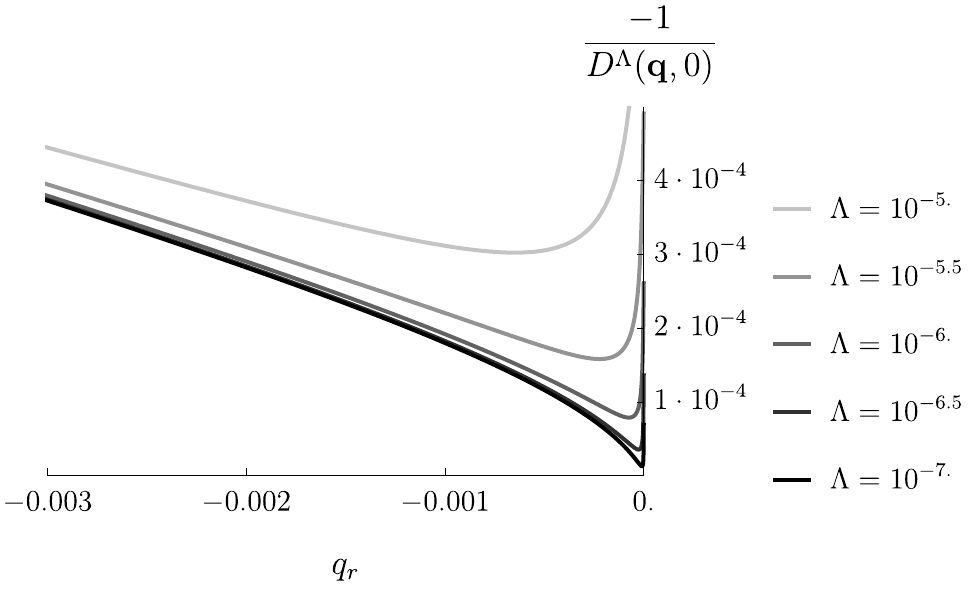}
\caption{Inverse fluctuation propagator $[D^{\Lam}(\bq,0)]^{-1}$ for $q_t = q_r$
 as a function of $q_r$ for small $q_r < 0$. The various curves have been obtained
 from a self-consistent solution for various choices of the infrared cutoff $\Lam$.}
\label{fig:D_sc}
\end{figure}
In Fig.~\ref{fig:D_sc} we show the momentum dependence of the inverse fluctuation propagator $[D^\Lam(\bq,0)]^{-1}$ for $q_t = q_r$ as a function of $q_r$, for various choices of the infrared cutoff $\Lam$. One can see that a well defined minimum is formed at $q_r = 0$ for $\Lam \to 0$. The point $q_r = q_t = 0$ is a minimum with respect to deviations in all directions. For $q_r = q_t > 0$ (not shown) the curves are much steeper than for $q_r = q_t < 0$. Hence, the $2k_F$ QCP with a nested ordering wave vector $\bQ$ is stabilized by self-consistently computed one-loop fluctuations. The slope of the inverse fluctuation propagator as a function of $q_r$ increases for small $q_r$ and small $\Lam$. The data in Fig.~\ref{fig:D_sc} suggest that the slope actually diverges for $q_r, \Lam \to 0$. In any case it is dominated by the hot spot contributions $\Pi_H$ and $\Pi_{H'}$ to the fluctuation propagator, while the regular contribution $b e_\bq + c q_t$ is comparatively small for small $q_r$. Hence, it looks as if the latter becomes irrelevant in the low energy, low momentum limit. The self-consistent solution thus seems to exhibit a higher degree of universality than the perturbative one-loop result, since the asymptotic behavior is determined entirely by the hot spot region, and the dependence on the parameters $b$ and $c$ disappears.



\section{Conclusion}

We have analyzed quantum fluctuation effects at the onset of charge or spin density wave order with an incommensurate $2k_F$ wave vector $\bQ$ in two-dimensional metals -- for the case where $\bQ$ connects two pairs of hot spots on the Fermi surface. This type of QCP is realized, for example, by the spin density wave instability of the Hubbard model at finite doping, \cite{schulz90,igoshev10} and by the onset of $d$-wave bond charge order generated by antiferromagnetic fluctuations in spin-fermion models for cuprates. \cite{metlitski10_af2,sachdev13,holder12}

We first computed the effective fluctuation propagator and the fermion self-energy at the QCP in a one-loop approximation without self-energy feedback. The marginal violation of Fermi liquid theory discovered in Ref.~\onlinecite{holder14} was thereby confirmed. As a function of the imaginary (Matsubara) frequency $k_0$, the real part of the self-energy at the hot spots is proportional to $|k_0|$ for small $k_0$, while the imaginary part is proportional to $|k_0| \log|k_0|$. This corresponds to an asymmetric linear frequency dependence of the imaginary part of the self-energy on the real frequency axis, with distinct coefficients for positive and negative frequencies. Unlike the case of the $2k_F$ QCP with a single hot spot pair, \cite{sykora18} the momentum dependence of the self-energy at the hot spots leads only to finite renormalizations of the Fermi velocity and the Fermi surface curvature. The Fermi velocity is enhanced by distinct renormalization factors for momenta inside and outside the Fermi sea, leading to a kink in the renormalized fermion dispersion at the hot spots.

Going beyond the leading order perturbation expansion we found that the one-loop result computed with bare propagators is not self-consistent. The particle-hole bubble with propagators dressed by the one-loop self-energy differs strongly from the bare particle-hole bubble. Taking only the singular frequency dependence of the self-energy into account, the peak of the particle-hole bubble at the $2k_F$ vector $\bQ$ is shifted away from the $2k_F$ line to a generic incommensurate wave vector. Hence, the QCP with a nested ordering wave vector, which is naturally favored in mean-field theory, seems to be spoiled by fluctuations. Including also the kink in the fermion dispersion generated by the momentum dependence of the self-energy, the peak remains pinned at the nesting vector at least for some choices of parameters, but its shape in the vicinity of $\bQ$ always differs qualitatively from the peak of the bare bubble.

We attempted to achieve self-consistency by performing a one-loop functional renormalization group calculation with self-energy feedback on the propagators. In this approach the ordering vector again turned out to be shifted away from the nesting vector $\bQ$, albeit only by a small amount. We believe that this result is an artifact of the extremely slow convergence of the fermion self-energy as a function of the flow parameter (an infrared frequency cutoff), which prevents a proper feedback of the self-energy in the flow equations.

Obtaining a fully self-consistent solution of the coupled one-loop integral equations with self-energy feedback and a high resolution at low energies is numerically challenging. Fortunately, the fRG calculation provided valuable hints on how to expand around the singular points, and how to use a slowly decreasing infrared cutoff as a technical device to converge to a self-consistent solution. In this solution the $2k_F$ QCP is not destroyed by fluctuations, at least for our choice of parameters. The peak in the fluctuation propagator (and the susceptibility) is even getting more pronounced, and its low energy structure is determined exclusively by low energy fluctuations in the hot spot region. While we cannot reach arbitrarily low frequencies, we get converged results over three frequency decades for the real part of the fermion self-energy, and five decades for the imaginary part (for imaginary frequencies). In that regime the real part is close to the result from one-loop perturbation theory, while the imaginary part follows the same $|k_0|\log|k_0|$ behavior, but with a significantly reduced prefactor. The marginal violation of Fermi liquid theory with a logarithmically vanishing quasi-particle weight obtained from the plain one-loop expansion is thus confirmed by the self-consistent calculation. The results on the imaginary frequency axis are consistent with a roughly linear frequency dependence of the imaginary part of the self-energy on the real frequency axis, with a steeper slope for negative (positive) frequencies, if the Fermi surface is convex (concave) at the hot spots.

In summary, we have established a new universality class for quantum critical behavior in two-dimensional metals -- at the onset of density wave order with an incommensurate nesting vector connecting two pairs of hot spots on the Fermi surface. It is worthwhile to further explore the critical behavior at and near this QCP. While we could reach rather low energy scales by our numerical solution of the self-consistent equations, one may try to extract the ultimate low energy behavior analytically. There are no relevant one-loop vertex corrections, but the role of higher order (two loop and beyond) corrections remains to be analyzed. One could also extend the analysis to the quantum critical regime at finite temperatures, look for secondary instabilities such as pairing, and study transport properties.


\begin{acknowledgments}
We are grateful to Pietro Bonetti, Lukas Debbeler, Johannes Mitscherling, and Demetrio Vilardi for valuable discussions.
\end{acknowledgments}

\vskip 5mm


\begin{appendix}

\section{Computation of one-loop self-energy}

In this appendix we derive the asymptotic results for the fermion self-energy, starting from the one-loop integral Eq.~(\ref{Sigma2}). To simplify the equations we set the global prefactor $M/N$ equal to one in the course of the derivation, and restore it in the results presented in the main text.


\subsection{Frequency dependence at hot spot}

For $\bk = \bk_H$, that is, at the hot spot, one has $k_r = k_t = 0$ and Eq.~(\ref{Sigma2}) reduces to
\begin{eqnarray} \label{SigmaA1}
 \Sg(\bk_H,ik_0) &=& \int \frac{dq_t}{2\pi} 
 \int \frac{dq_r}{2\pi} \int \frac{dq_0}{2\pi} \, 
 \frac{1}{i(k_0-q_0) - v_F q_r - \frac{1}{2m} q_t^2} \nonumber \\
 & \times & \frac{1}
 {a h(e_\bq,q_0) + a' h(e'_\bq,q_0) - b e_\bq - c q_t} \, ,
\end{eqnarray}
The form of the fermion propagator indicates the following scaling of the integration variables in the low-frequency limit $k_0 \to 0$:
\begin{equation}
 q_0 \sim k_0 \, , \; q_r \sim k_0 \, , \; q_t \sim \sqrt{|k_0|} \, .
\end{equation}
Hence, $q_r/q_t \ll 1$, such that $e'_\bq \approx v'_F \sin\theta \, q_t$.
The contributions to the denominator of the effective interaction in Eq.~(\ref{SigmaA1}) thus scale as follows:
$ah(e_\bq,q_0) \sim \sqrt{|k_0|}$, $b e_\bq \sim k_0$, $c q_t \sim \sqrt{|k_0|}$,
$a' h(e'_\bq,q_0) \sim |k_0|^{1/4}$ for $\sin\theta \, q_t > 0$ and
$a' h(e'_\bq,q_0) \sim |k_0|^{3/4}$ for $\sin\theta \, q_t < 0$.
Hence, contributions from $\sin\theta \, q_t > 0$ are subleading compared to the contributions from $\sin\theta \, q_t < 0$. In the latter region the largest terms in the denominator are of order $\sqrt{|k_0|}$, and the terms of higher order in $k_0$ are negligible.
Assuming $\theta > 0$ for definiteness, and keeping only the leading terms, the self-energy can be written as
\begin{eqnarray} \label{SigmaA2}
 \Sg(\bk_H,ik_0) &=& \int_{q_t < 0} \! \frac{dq_t}{2\pi} 
 \int \frac{dq_r}{2\pi} \int \frac{dq_0}{2\pi} \, 
 \frac{1}{i(k_0-q_0) - e_\bq - \frac{1}{4m} q_t^2} \nonumber \\
 & \times & \frac{1}
 {a h(e_\bq,q_0) - c q_t} \, .
\end{eqnarray}
Note that we have expressed the kinetic energy in the fermion propagator in terms of $e_\bq$ and $q_t$ instead of $q_r$ and $q_t$. It is now convenient to use $e_\bq$ as integration variable instead of $q_r$. The Jacobian for this substitution is $1/v_F$.

Introducing dimensionless variables via $q_0 = |k_0| \tilde q_0$,
$e_\bq = |k_0| \tilde e_\bq$, and $q_t = \sqrt{4m|k_0|} \tilde q_t$, and using
$a = \sqrt{m}/(4\pi v_F)$ one obtains
\begin{equation} \label{SigmaA3}
 \Sg(\bk_H,ik_0) = \frac{|k_0|}{\pi^2} 
 \int_{\tilde q_t < 0} \hskip -4mm d \tilde q_t
 \int \! d \tilde e_\bq 
 \int \! d \tilde q_0 \,
 \frac{1}{i(s_0 - \tilde q_0) - (\tilde e_\bq + \tilde q_t^2)} \,
 \frac{1}{\sqrt{2} \big[(\tilde e_\bq^2 + \tilde q_0^2)^\frac{1}{2} + 
 \tilde e_\bq \big]^\frac{1}{2} - \alf \tilde q_t} \, ,
\end{equation}
where $s_0 = \sgn(k_0)$. The integrand depends only on a single dimensionless parameter,
\begin{equation} \label{alpha}
 \alf = 8\pi v_F \, c \, .
\end{equation}
We recall that $c$ needs to be positive to have a peak in the susceptibility and effective interaction, if $\theta > 0$. For $\theta < 0$, $c$ needs to be negative and the self-energy has the same form with $\alf = - 8\pi v_F \, c = 8\pi v_F |c|$.


\subsubsection{Imaginary part}

The imaginary part of the integral in Eq.~(\ref{SigmaA3}) diverges logarithmically. The divergence is due to contibutions from the regime $\tilde e_\bq < 0$ with
$|\tilde e_\bq| \gg |\tilde q_0|$, where
$\sqrt{2} \big[(\tilde e_\bq^2 + \tilde q_0^2)^\frac{1}{2} +
 \tilde e_\bq \big]^\frac{1}{2} \approx |\tilde q_0|/|\tilde e_\bq|^\frac{1}{2}$.
The leading contribution to the imaginary part of the self-energy can thus be written as
\begin{equation} \label{ImSigmaA1}
 \Im\Sg(\bk_H,ik_0) = \frac{|k_0|}{\pi^2} 
 \int_{\tilde q_t < 0} \hskip -4mm d \tilde q_t
 \int_{\tilde e_\bq < 0} \hskip -4mm d \tilde e_\bq 
 \int \! d \tilde q_0 \
 \frac{\tilde q_0 - s_0}{(\tilde q_0 - s_0)^2 + (\tilde e_\bq + \tilde q_t^2)^2} \,
 \frac{1}{|\tilde q_0|/\sqrt{|\tilde e_\bq|} - \alf \tilde q_t} \, .
\end{equation}
Introducing new momentum space variables
$x = \alf \sqrt{|\tilde e_\bq| \tilde q_t^2}$ and
$y = \tilde e_\bq + \tilde q_t^2$, and using 
\begin{equation}
 \sqrt{|\tilde e_\bq}| \, d \tilde e_\bq \, \tilde dq_t =
 \frac{1}{2\alf} \left( 1 - y/\sqrt{y^2 + 4x^2/\alf^2} \right) dxdy \, ,
\end{equation}
one finds
\begin{equation}
 \Im\Sg(\bk_H,ik_0) = \frac{|k_0|}{2\pi^2 \alf}
 \int_{x>0} \hskip -3mm dx \int dy \int d \tilde q_0 \,
 \Big( 1 - \frac{y}{\sqrt{y^2 + 4x^2/\alf^2}} \Big) \,
 \frac{\tilde q_0 - s_0}{(\tilde q_0 - s_0)^2 + y^2} \,
 \frac{1}{|\tilde q_0| + x} \, .
\end{equation}
The second term in the bracket is odd in $y$ and therefore yields no contribution to the integral. The remaining integral is elementary. The integrations over $y$ and $\tilde q_0$ are convergent (without any UV cutoff), yielding
\begin{eqnarray}
 \Im\Sg(\bk_H,ik_0) &=& \frac{|k_0|}{2\pi^2 \alf}
 \int_{x>0} \hskip -3mm dx \int d \tilde q_0 \int dy \,
 \frac{\tilde q_0 - s_0}{(\tilde q_0 - s_0)^2 + y^2} \,
 \frac{1}{|\tilde q_0| + x} \nonumber \\[1mm]
 &=& \frac{|k_0|}{2\pi \alf}
 \int_{x>0} \hskip -3mm dx \int d \tilde q_0 \,
 \frac{\sgn(\tilde q_0 - s_0)}{|\tilde q_0| + x} \, = \, 
 - \frac{k_0}{\pi \alf} \int_{x>0} \hskip -3mm dx \,
 \ln \big( 1 + x^{-1} \big) \, . \hskip 1cm
\end{eqnarray}
The $x$-integral diverges logarithmically for large $x$.
Restricting $[|e_\bq| q_t^2/(4m)]^\frac{1}{2}$ by a UV cutoff $\Lam$ leads to a cutoff $\alf\Lam/|k_0|$ for the variable $x$. With this cutoff, the $x$-integral yields
\begin{equation}
 \Im\Sg(\bk_H,ik_0) =
 - \frac{k_0}{\pi \alf} \, \ln \frac{\alf\Lam}{|k_0|} + {\cal O}(k_0)
\end{equation}
for $k_0 \to 0$.


\subsubsection{Real part}

To evaluate the real part of the self-energy at the hot spot, $\Re\Sg(\bk_H,ik_0)$, we start from the expression (\ref{SigmaA3}).
The integral for the real part diverges linearly in the ultraviolet regime. To obtain a finite result one needs to subtract the self-energy at $k_0 = 0$, which is given by the same expression with $s_0$ replaced by zero.

The $\tilde q_t$ integration can be carried out analytically by a keyhole contour integration around the negative real axis. To this end, we rewrite Eq.~(\ref{SigmaA3}) in the form
\begin{equation} \label{ReSigmaA1}
 \Sg(\bk_H,ik_0) = \frac{|k_0|}{\pi^2 \alf} 
 \int d \tilde e_\bq \int d \tilde q_0 \int_{-\infty}^0 \!\! d \tilde q_t \;
 \frac{1}{\tilde q_t^2 - A^2} \, \frac{1}{\tilde q_t - B} \, ,
\end{equation}
where
\begin{equation}
 A = \sqrt{- \tilde e_\bq + i(s_0 - \tilde q_0)} \, , \quad
 B = \frac{\sqrt{2}}{\alf} 
     \sqrt{\sqrt{\tilde e_\bq^2 + \tilde q_0^2} + \tilde e_\bq} \, .
\end{equation}
The contour integration yields
\begin{equation} \label{int_qt}
 \int_{-\infty}^0 \!\! d \tilde q_t \;
 \frac{1}{\tilde q_t^2 - A^2} \, \frac{1}{q_t - B} =
 \frac{\ln(A)}{2A(A-B)} + \frac{\ln(-A)}{2A(A+B)} - \frac{\ln(B)}{A^2-B^2} \, .
\end{equation}
The real part of the self-energy at the hot spot can thus be written as
\begin{equation}
 \Re\Sg(\bk_H,ik_0) = - \frac{|k_0|}{\pi\alf} R(\alf) \, ,
\end{equation}
where the function $R(\alf)$ is given by the double integral
\begin{equation} \label{R_alf}
 R(\alf) = - \frac{1}{\pi} \int d \tilde e_\bq \int d \tilde q_0 \, \Re \left[ 
 \frac{\ln(A)}{2A(A-B)} + \frac{\ln(-A)}{2A(A+B)} - \frac{\ln(B)}{A^2-B^2}
 \right] \, .
\end{equation}

The remaining two integrations seem to be difficult to handle analytically. However, a simple analytic result can be obtained for the real part in the limit of large $\alf$. In this limit, $B$ in Eq.~(\ref{int_qt}) can be set to zero except in the argument of the logarithm in the last term.

For $B=0$, the real parts of the first two terms in Eq.~(\ref{int_qt}) integrate to zero. To see this, we temporarily introduce ultraviolet cutoffs $\pm \tilde\Lam$ for the integration variables $\tilde e_\bq$ and $\tilde q_0$.
The $\tilde e_\bq$ integration of the first term yields
\begin{eqnarray}
 I_1(s_0) &=& \frac{1}{2} 
 \int_{-\tilde\Lam}^{\tilde\Lam} \!\! d \tilde q_0
 \int_{-\tilde\Lam}^{\tilde\Lam} \!\! d \tilde e_\bq \,
 \Re \frac{\ln \big[\sqrt{- \tilde e_\bq + i(s_0 - \tilde q_0)} \, \big]}
 {- \tilde e_\bq + i(s_0 - \tilde q_0)} \nonumber \\[2mm]
 &=& \frac{\pi}{8}
 \int_{-\tilde\Lam}^{\tilde\Lam} \!\! d \tilde q_0
 \int_{-\tilde\Lam}^{\tilde\Lam} \!\! d \tilde e_\bq \,
 \frac{|\tilde q_0 - s_0|}{\tilde e_\bq^2 + (\tilde q_0 - s_0)^2} =
 \frac{\pi}{4} \int_{-\tilde\Lam}^{\tilde\Lam} \!\! d \tilde q_0
 \arctan \frac{\tilde\Lam}{|\tilde q_0 - s_0|} \, .
\end{eqnarray}
Obviously $I_1(s_0)$ diverges linearly for $\tilde\Lam \to \infty$. However, this divergence is canceled upon subtracting $I_1(0)$, corresponding to the subtraction of the self-energy at $k_0 = 0$, and the difference $I_1(s_0) - I_1(0)$ even vanishes in the limit $\tilde\Lam \to \infty$. For example, for $s_0 = 1$ one finds
\begin{equation}
 I_1(1) - I_1(0) =
 \frac{\pi}{8} \int_{\tilde\Lam}^{\tilde\Lam + 1} d \tilde q_0 \,
 \arctan \frac{\tilde\Lam}{|\tilde q_0|} -
 \frac{\pi}{8} \int_{\tilde\Lam - 1}^{\tilde\Lam} d \tilde q_0 \,
 \arctan \frac{\tilde\Lam}{|\tilde q_0|} \sim \tilde\Lam^{-1} \to 0 \, .
\end{equation}
By the same reasoning, the second contribution $I_2(s_0) - I_2(0)$ vanishes, too.

The real part of the third contribution has the form
\begin{eqnarray}
 I_3(s_0) &=& 
 \int_{-\infty}^{\infty} \!\! d \tilde q_0 
 \int_{-\infty}^{\infty} \!\!  d \tilde e_\bq \,
 \ln \left( \frac{\sqrt{2}}{\alf}
 \sqrt{\sqrt{\tilde e_\bq^2 + \tilde q_0^2} + \tilde e_\bq} \right) \,
 \Re \frac{1}{\tilde e_\bq - i(s_0 - \tilde q_0)} \\
 &=& \frac{1}{2}
 \int_0^{\infty} \!\! d \tilde q_0 
 \int_0^{\infty} \!\!  d \tilde e_\bq \,
 \ln \left( \frac{\sqrt{\tilde e_\bq^2 + \tilde q_0^2} + \tilde e_\bq}
 {\sqrt{\tilde e_\bq^2 + \tilde q_0^2} - \tilde e_\bq} \right)
 \left( \frac{\tilde e_\bq}{\tilde e_\bq^2 + (\tilde q_0 - s_0)^2} +
 \frac{\tilde e_\bq}{\tilde e_\bq^2 + (\tilde q_0 + s_0)^2} \right)
 \nonumber
\end{eqnarray}
for large $\alf$.
Substituting $\tilde q_0$ by $u = \tilde q_0/\tilde e_\bq$, and subtracting the zero frequency constant yields
\begin{eqnarray}
 I_3(s_0) - I_3(0) &=& \frac{1}{2} \int_0^{\infty} \!\! du \,
 \ln \left( \frac{\sqrt{1+u^2} + 1}{\sqrt{1+u^2} - 1} \right) \nonumber \\
 &&\times \int_0^{\infty} \!\! d \tilde e_\bq \left(
 \frac{\tilde e_\bq^2}{\tilde e_\bq^2 + (u \tilde e_\bq - 1)^2} +
 \frac{\tilde e_\bq^2}{\tilde e_\bq^2 + (u \tilde e_\bq + 1)^2} -
 \frac{2}{1 + u^2} \right) \, . \hskip 1cm
\end{eqnarray}
The integral over $\tilde e_\bq$ can be extended to the entire real axis (with a compensation by a factor~$\frac{1}{2}$), since the integrand is symmetric in $\tilde e_\bq$. The $\tilde e_\bq$ integration can then be easily evaluated by the residue theorem, yielding
\begin{equation}
 \int_0^{\infty} \!\! d \tilde e_\bq \left(
 \frac{\tilde e_\bq^2}{\tilde e_\bq^2 + (u \tilde e_\bq - 1)^2} +
 \frac{\tilde e_\bq^2}{\tilde e_\bq^2 + (u \tilde e_\bq + 1)^2} -
 \frac{2}{1 + u^2} \right) =
 \pi \, \frac{u^2 - 1}{(u^2 + 1)^2} \, .
\end{equation}
The $u$-integral can now be performed using an integration by parts, yielding
\begin{equation}
 I_3(s_0) - I_3(0) = - \pi \, .
\end{equation}
Inserting this into Eq.~(\ref{ReSigmaA1}) we obtain
\begin{equation}
 \Re\Sg(\bk_H,ik_0) - \Sg(\bk_H,0) =
 - \frac{|k_0|}{\pi \alf}
\end{equation}
for large $\alf$.


\subsection{Momentum dependence near hot spot}

We now evaluate the momentum dependence of the self-energy near the hot spot at zero frequency. The momentum is parametrized by the normal and tangential coordinates relative to the hot spot, $k_r$ and $k_t$, respectively.

We first analyze the {\em normal}\/ momentum dependence for $k_t = 0$, which we denote by $\Sg(k_r,0)$, where the frequency argument has been dropped. Eq.~(\ref{Sigma2}) yields
\begin{eqnarray} \label{SigmakA1}
 \Sg(k_r,0) &=& \int \frac{dq_t}{2\pi} 
 \int \frac{dq_r}{2\pi} \int \frac{dq_0}{2\pi} \, 
 \frac{1}{-i q_0 - v_F (q_r - k_r) - \frac{1}{2m} q_t^2} \nonumber \\
 & \times & \frac{1}
 {a h(e_\bq,q_0) + a' h(e'_\bq,q_0) - b e_\bq - c q_t } \, .
\end{eqnarray}
The form of the fermion propagator indicates the following scaling of the integration variables in the limit $k_r \to 0$:
\begin{equation}
 q_0 \sim k_r \, , \; q_r \sim k_r \, , \; q_t \sim \sqrt{|k_r|} \, .
\end{equation}
Following the same arguments as for the frequency dependence, the expression (\ref{SigmakA1}) can be approximated by
\begin{equation} \label{SigmakA2}
 \Sg(k_r,0) = \frac{v_F|k_r|}{\pi^2} 
 \int_{\tilde q_t < 0} \hskip -4mm d \tilde q_t
 \int \! d \tilde e_\bq 
 \int \! d \tilde q_0 \,
 \frac{1}{s_{k_r} -i\tilde q_0 - (\tilde e_\bq + \tilde q_t^2)} \,
 \frac{1}{\sqrt{2} \big[(\tilde e_\bq^2 + \tilde q_0^2)^\frac{1}{2} + 
 \tilde e_\bq \big]^\frac{1}{2} - \alf \tilde q_t} \, ,
\end{equation}
for small $k_r$, where $s_{k_r} = \sgn(k_r)$, and $\alf$ is the same as in Eq.~(\ref{alpha}). The dimensionless integration variables are defined by $q_0 = v_F|k_r| \, \tilde q_0$,
$e_\bq = v_F |k_r| \tilde e_\bq$, and $q_t = \sqrt{4mv_F|k_r|} \, \tilde q_t$.
The $\tilde q_t$ integral can now be performed by a keyhole contour integration around the negative real axis, yielding
\begin{equation} \label{SigmakA3}
 \Sg(k_r,0) = \frac{v_F |k_r|}{\pi^2 \alf}
 \int \! d \tilde e_\bq \int \! d \tilde q_0 \left(
 \frac{\ln(A)}{2A(A-B)} + \frac{\ln(-A)}{2A(A+B)} - \frac{\ln(B)}{A^2-B^2}
 \right) ,
\end{equation}
where
\begin{equation}
 A = \sqrt{s_k - \tilde e_\bq - i \tilde q_0} \, , \quad
 B = \frac{\sqrt{2}}{\alf} 
     \sqrt{\sqrt{\tilde e_\bq^2 + \tilde q_0^2} + \tilde e_\bq} \, .
\end{equation}
After subtracting the constant $\Sigma(0,0)$, the remaining integrations are convergent, that is, no ultraviolet cutoff is needed. One thus obtains
\begin{equation}
 \Sg(k_r,0) - \Sg(0,0) = a_{r,s_{k_r}}(\alf) \, v_F k_r \, ,
\end{equation}
where $a_{r,\pm}(\alf)$ are positive numbers depending only on $\alf$ and the sign of $k_r$. An analytic evaluation of the remaining two integrations in Eq.~(\ref{SigmakA3}) seems difficult, but a numerical evaluation can be done with high accuracy.

We now turn to the {\em tangential}\/ momentum dependence of the self-energy for $k_r = 0$. Eq.~(\ref{Sigma2}) yields
\begin{eqnarray} \label{SigmakA4}
 \Sg(0,k_t) &=& \int \frac{dq_t}{2\pi} 
 \int \frac{dq_r}{2\pi} \int \frac{dq_0}{2\pi} \, 
 \frac{1}{-i q_0 - v_F q_r - \frac{1}{2m} (q_t - k_t)^2} \nonumber \\
 & \times & \frac{1}
 {a h(e_\bq,q_0) + a' h(e'_\bq,q_0) - b e_\bq - c q_t} \, .
\end{eqnarray}
The form of the fermion propagator indicates the following scaling of the integration variables in the limit $k_t \to 0$:
\begin{equation}
 q_0 \sim k_t^2 \, , \; q_r \sim k_t^2 \, , \; q_t \sim k_t \, .
\end{equation}
Following the same arguments as for the frequency dependence, the expression (\ref{SigmakA4}) can be approximated by
\begin{equation} \label{SigmakA5}
 \Sg(0,k_t) = - \frac{k_t^2}{4\pi^2 m} 
 \int_{\tilde q_t < 0} \hskip -4mm d \tilde q_t
 \int \! d \tilde e_\bq 
 \int \! d \tilde q_0 \,
 \frac{1}{i\tilde q_0 + \tilde e_\bq + (\tilde q_t - s_t)^2} \,
 \frac{1}{\sqrt{2} \big[(\tilde e_\bq^2 + \tilde q_0^2)^\frac{1}{2} + 
 \tilde e_\bq \big]^\frac{1}{2} - \alf \tilde q_t} \, ,
\end{equation}
for small $k_t$, where $s_t = \sgn(k_t)$, and $\alf$ is the same as in Eq.~(\ref{alpha}). The dimensionless integration variables are defined by
$q_0 = \frac{k_t^2}{4m} \, \tilde q_0$,
$e_\bq = \frac{k_t^2}{4m} \tilde e_\bq$, and $q_t = |k_t| \, \tilde q_t$.
The $\tilde q_t$ integral can again be performed by a keyhole contour integration around the negative real axis, yielding
\begin{equation} \label{SigmakA6}
 \Sg(0,k_t) = \frac{k_t^2}{4\pi^2 m \alf}
 \int \! d \tilde e_\bq \int \! d \tilde q_0 \left(
 \frac{\ln(s_t + A)}{2A(s_t+A-B)} + \frac{\ln(s_t-A)}{2A(A+B-s_t)} +
 \frac{\ln(B)}{(s_t-B)^2-A^2}
 \right) ,
\end{equation}
where
\begin{equation}
 A = \sqrt{- \tilde e_\bq - i \tilde q_0} \, , \quad
 B = \frac{\sqrt{2}}{\alf} 
     \sqrt{\sqrt{\tilde e_\bq^2 + \tilde q_0^2} + \tilde e_\bq} \, .
\end{equation}
The integral in Eq.~(\ref{SigmakA6}) remains ultraviolet divergent even after subtracting $\Sg(0,0)$. Restricting $e_\bq$ and $q_0$ by an ultraviolet cutoff $\Lam$, a numerical evaluation of the integals yields
\begin{equation}
 \Sg(0,k_t) - \Sg(0,0) = a_t(\alf) \sqrt{\Lam/m} \, k_t + {\cal O}(k_t^2) \, ,
\end{equation}
where $a_t(\alf) \sim \log(\alf)$ for small $\alf$ and $a_t(\alf) \sim \log(\alf)/\alf$ for large $\alf$.
Note that a cutoff $\Lam$ on the variables $e_\bq$ and $q_0$ entails a cutoff $\tilde\Lam = \frac{4m}{k_t^2} \Lam$ on the rescaled variables $\tilde e_\bq$ and $\tilde q_0$. The square-root-type UV divergence of the integral in Eq.~(\ref{SigmakA6}) thus yields a factor proportional to $\sqrt{\Lam}/k_t$, which reduces the nominally quadratic $k_t$ dependence of $\Sg(0,k_t) - \Sg(0,0)$ to a linear dependence.


\section{Computation of particle-hole bubble}

In the following we evaluate the particle-hole bubble $\Pi(\bq,iq_0)$ defined in Eq.~(\ref{Pi}) with a propagator of the form (\ref{G}). Using radial and tangential momentum coordinates and expanding the fermion dispersion yields 
\begin{equation}
 \Pi(\bq,iq_0) = N \int \frac{d^2\bk}{(2\pi)^2} \int \frac{dk_0}{2\pi} \,
 \frac{1}{\{k_0\} - v_F k_r - \frac{k_t^2}{2m}} \,
 \frac{1}{\{k_0 - q_0\} + v_F(k_r - q_r) - \frac{(k_t - q_t)^2}{2m}} \, .
\end{equation}
Here we have introduced the short-hand notation $\{k_0\} = ik_0 - \Sg_H(ik_0)$, where $\Sg_H(ik_0) = \Sg(\bk_H,ik_0) - \Sg(\bk_H,0)$.
The $k_r$ and $k_t$ integrations can be performed via the residue theorem,
\begin{eqnarray} \label{PiA1}
 \Pi(\bq,iq_0) &=& N \frac{i}{v_F} \int \frac{dk_0}{2\pi} \frac{dk_t}{2\pi} \,
 \frac{\Theta(-k_0) - \Theta(k_0 - q_0)}
 {\{k_0\} + \{k_0 - q_0\} - v_F q_r - \frac{k_t^2}{2m} - \frac{(k_t - q_t)^2}{2m}}
 \nonumber \\[2mm]
 &=& N \frac{i}{v_F} \int \frac{dk_0}{2\pi} \int \frac{dk_t}{2\pi} \,
 \frac{\Theta(k_0 - q_0) - \Theta(-k_0)}
 {\frac{(k_t - q_t/2)^2}{m} + e_\bq - \{k_0\} + \{k_0 - q_0\}}
 \nonumber \\[2mm]
 &=& - N \frac{\sqrt{m}}{4\pi v_F} \int dk_0 \,
 \frac{|\Theta(k_0 - q_0) - \Theta(-k_0)|}
 {\sqrt{-e_\bq + \{k_0\} + \{k_0 - q_0\}}} \, .
\end{eqnarray}
Subtracting $\Pi(\bQ,0)$ one thus obtains
\begin{equation} \label{deltaPiA1}
 \delta\Pi(\bq,iq_0) = - N \frac{\sqrt{m}}{4\pi v_F} \left[
 \int dk_0 \, \frac{|\Theta(k_0 - q_0) - \Theta(-k_0)|}
 {\sqrt{-e_\bq + \{k_0\} + \{k_0 - q_0\}}} -
 \int \frac{dk_0}{\sqrt{2\{k_0\}}} \right] \, .
\end{equation}
The remaining integral can be easily carried out numerically.

For the bare bubble one obtains the same expression with $\{k_0\} = ik_0$. The $k_0$-integration is then convergent (without UV cutoff) and elementary, yielding the familiar singular contribution
\begin{equation}
 \delta\Pi_0(\bq,iq_0) = N \frac{\sqrt{m}}{4\pi v_F}
 \left( \sqrt{e_\bq + iq_0} + \sqrt{e_\bq - iq_0} \right) \, .
\end{equation}

\end{appendix}



\begin{thebibliography}{99}
\bibitem{footnote_incomm} Commensurate wave vectors can be written as a
 linear combination of reciprocal lattice vectors with rational coefficients,
 while incommensurate wave vectors cannot. Obviously there are infinitly
 many possible rational coefficients, but the quantitative effects of
 commensurability decrease rapidly with the size of their denominators.
\bibitem{footnote_chi_0} In Ref.~\onlinecite{holder14} the regular contributions to $\chi_0$ were parametrized by a sum of four terms involving normal and tangential coordinates of both $2k_F$ lines. Two of the terms are however redundant, since two parameters are sufficient to capture the leading (linear) momentum dependence.
\bibitem{footnote3}
As illustrative examples, the structure of the effective interaction is derived for a spinless fermion model and for the Hubbard model in Ref.~\onlinecite{sykora18}.
\bibitem{footnote4}
Our parameter $\alf$ corresponds to $2 \tilde b$ in Ref.~\onlinecite{holder14}, and our angle $\theta$ corresponds to the angle $-\phi$. Moreover, $\phi$ was tacidly assumed to be positive in Ref.~\onlinecite{holder14}.
\bibitem{footnote_Z}
Our definition of $Z$ follows the usual convention in quantum field theory. In textbooks on Fermi liquid theory, however, the quasi-particle weight itself (not its inverse) is frequently denoted by $Z$.
\bibitem{footnote_num} In the numerical solution of the flow equations from Sec.~V,
 and of the self-consistent integral equations from Sec.~VI, we actally parametrize the
 regular contributions to the inverse fluctuation propagator as $b \, e_\bq + b' e'_\bq$,
 and choose $b = b' = 1/(8\pi)$. For $\theta = \pi/2$, this is equivalent to
 $b \, e_\bq + c \, q_t$ with $b = c = 1/(8\pi)$ up to subleading corrections, since
 $e'_\bq = q_t + q_r^2/4$ for $\theta = \pi/2$ and $v_F = m = 1$.
\end{thebibliography}
\end{document}